\documentclass[12pt]{iopart}

\usepackage{iopams}
\usepackage{graphicx}
\usepackage{amssymb}
\usepackage{bm}
\usepackage{pifont}
\usepackage{color}
\usepackage[FIGTOPCAP]{subfigure}
\usepackage[latin1]{inputenc}

\newcommand{\oma}{\mbox{\boldmath $\Omega_1$}}
\newcommand{\omb}{\mbox{\boldmath $\Omega_2$}}
\newcommand{\omc}{\mbox{\boldmath $\Omega_3$}}
\newcommand{\omd}{\mbox{\boldmath $\Omega_4$}}
\newcommand{\ome}{\mbox{\boldmath $\Omega_5$}}
\newcommand{\omi}{\mbox{\boldmath $\Omega_i$}}
\newcommand{\sig}{\mbox{\boldmath $\sigma$}}
\newcommand{\siga}{\mbox{\boldmath $\sigma_1$}}

\newcommand{\HT}{\mbox{\boldmath $H$}}
\newcommand{\QTa}{\mbox{\boldmath $Q_1$}}
\newcommand{\QTb}{\mbox{\boldmath $Q_2$}}
\newcommand{\UN}{\mbox{\boldmath $I$}}

\newcommand{\TX}{\mbox{\boldmath $\hat{t}$}}
\newcommand{\NX}{\mbox{\boldmath $\hat{n}$}}

\def\3{\ss }           % scharf-ss
\begin{document}

\title{Chapman-Enskog expansion for the Vicsek model of self-propelled particles} 
\author{Thomas Ihle}
\address{Institut f{\"u}r Physik,
Ernst-Moritz-Arndt-Universit{\"a}t Greifswald, 17489 Greifswald, Germany}
\address{Department of Physics, North Dakota State University, Fargo, ND 58108-6050, USA}
\ead{thomas.ihle@uni-greifswald.de}

\begin{abstract}
Using the standard Vicsek model, 
I show how the macroscopic transport equations can be systematically derived from microscopic collision rules.
The approach starts with the exact evolution equation for the N-particle probability distribution, and after
making
the mean-field assumption of Molecular Chaos 
leads to a multi-particle Enskog-type equation.
This equation is treated by a non-standard Chapman-Enskog expansion 
to extract the macroscopic behavior. 
The expansion includes terms up to third order in a formal expansion parameter $\epsilon$, and involves a fast time scale. 
A self-consistent closure of the moment equations is presented that leads to a continuity equation for the particle density and a Navier-Stokes-like equation for the momentum density. Expressions for 
all transport coefficients in these macroscopic equations
are given explicitly in terms of microscopic parameters of the model. 
The transport coefficients depend on specific angular integrals which are evaluated asymptotically in the limit of
infinitely many collision partners, using an analogy to a random walk.
The consistency of the Chapman-Enskog approach is checked by an independent calculation of 
the shear viscosity using 
a Green-Kubo relation.
\end{abstract}


\maketitle

%\ioptwocol

\section{Introduction}

Recently, there has been a large interest in active many-particle systems such as 
bird flocks \cite{animal_flocks}, swarming bacteria \cite{BenJacob_97},
chemically powered nanorods \cite{nano_rods}, 
microtubule mixtures 
\cite{nedelec_02} and 
actin networks \cite{actin_net} driven by molecular motors. 
These systems display very interesting behaviors such as pattern formation, collective motion and non-equilibrium phase 
transitions \cite{vicsek_12,marchetti_13}. 
Due to the many degrees of freedom, theoretical and numerical investigations are usually based on  
coarse-grained macroscopic transport equations for the slow variables.   
In many cases, the general form of these equations are designed by symmetry arguments or Onsager relations which 
means that the exact coefficients of the terms in them remain undetermined, see for example the 
Toner-Tu theory \cite{toner_95,toner_98}.
This can lead to models with many free parameters which are time-consuming to explore numerically. 
In addition, in complicated systems it is possible to overlook nontrivial symmetries which would cause some of these terms to vanish.
It is also not clear how these many terms depend on a few microscopic parameters, and brute force
scanning of the parameters could waste computer time in regions of the parameter space which are not compatible with the underlying
microscopic system. 

Therefore, it is very desirable to have a direct, rigorous derivation of these macroscopic equations from
the microscopic rules.
Systems of interest 
are often 
described by simplified models, such as the Vicsek-model (VM) for self-propelled agents, 
\cite{vicsek_95,czirok_97,nagy_07},  which are suited for computer simulations but keep the essential physics. 
The time evolution in such models is often discrete, that is, there is a non-zero time step, $\tau$, and 
there is some type of generalized interaction between the simulated objects.
For instance, in the VM, at every time step an agent interacts with all agents in a circle of radius $r_0$ around it
to adopt its flying direction towards the mean direction of the other agents plus some noise.
Interpreting agents as ``particles'', this alignment corresponds to a genuine multi-particle interaction which
is not pairwise additive.

In this paper, I will present how
to systematically derive macroscopic evolution equations from the microscopic rules 
of the standard Vicsek-model using techniques from kinetic theory.
One of the techniques applied in this paper is a non-standard Chapman-Enskog expansion that keeps a fast time scale.
This approach can be easily generalized to other particle-based models of active soft matter \cite{marchetti_13} 
with a discrete time step and multi-particle interactions.
In previous publications \cite{ihle_11,ihle_13,ihle_14_a,ihle_15_a}, 
the results of this derivation have already been discussed and used without showing the technical details.
For example, in table 2 of Ref. \cite{ihle_15_a} the density dependence of the transport coefficients 
in the large density limit was given.
In this paper, I will focus on the details of the derivation.
This includes presenting a self-consistent closure of the moment equations as well as determining the coefficients of nonlinear terms and gradient terms in the hydrodynamic equations up to a predefined order.

Note, that similar derivations of macroscopic equations for Vicsek-like models and similar closures have been performed 
by other groups, see for example 
Refs. \cite{bertin_06,bertin_09,baskaran_08b,grossmann_13,chepizhko_14,peshkov_12,farrell_12,peshkov_14_b}.
However, these authors did not explicitly 
treat the standard VM with multi-body interactions and discrete time step as done in this 
paper. Furthermore, they used methods to extract the macroscopic behavior, 
which differ from the ones used here.
Recently, the results presented in this paper
and the ones obtained by the Boltzmann-approach of Refs. \cite{bertin_06,bertin_09} were compared in detail 
and critically debated in a series of publications, Refs. \cite{ihle_14_a,peshkov_14_b,ihle_14_b,bertin_14_a,ihle_14_c}.
However, a comprehensive comparative study of the different approaches combined with computer-assisted 
quantitative verification of the derived 
transport coefficients is still lacking and will be left for the future.

\section{The Vicsek model}
\label{sec:Vicsek}
The two-dimensional Vicsek-model (VM) \cite{vicsek_95,czirok_97,nagy_07} 
consists of $N$ point particles at global number density $\rho_0$, which move at constant
speed $v_0$.
The positions and velocities of the particles are given by 
${\bf x}_i(t)$ and ${\bf v}_i(t)$, respectively.
The particles
undergo discrete-time dynamics
with time step $\tau$. The evolution consists of two steps: streaming and collision.
Note, that the term ``collision'' is not to be taken literally. Instead, it just denotes any interaction that
changes the momentum of a particle.
In the streaming step
all positions are updated according to
\begin{equation}
\label{STREAM}
{\bf x}_i(t+\tau)={\bf x}_i(t)+\tau {\bf v}_i(t)\,.
\end{equation}
Because the particle speeds stay the same at all times, the velocities are parameterized by the ``flying'' angles, $\theta_i$,
${\bf v}_i=v_0(\cos{\theta_i},\sin{\theta_i})$.
In the collision step,
the directions $\theta_i$ 
are modified.
Particles align with their neighbors within a fixed distance $r_0$ plus some external noise:
a circle of radius $r_0$ is drawn around the focal particle $i$, and the average direction $\Phi_i$ of motion
of the
particles (including particle $i$)
within the circle is determined
according to
\begin{equation}
\label{COLLIS}
\Phi_i={\rm Arg}\left( \sum_{\{j\}} {\rm e}^{i\theta_j}
\right)
%{\rm arctan}[\sum_{\{j\}} {\rm sin}(\theta_j)/\sum_{\{j\}} {\rm cos}(\theta_j)]\,,
\end{equation}
Eq. (\ref{COLLIS}) means that the vector sum of all particle velocities in every circle is computed and the direction of the resulting 
vector
is taken as average angle $\Phi_i$.
Once all average directions $\Phi_i$ are known, the new directions
follow as
\begin{equation}
\label{VM_RULE}
\theta_i(t+\tau)=\Phi_i+\xi_i
\end{equation}
 where $\xi_i$ is the so-called angular noise. The random numbers $\xi_i$ are
uniformly distributed 
in the interval $[-\eta/2,\eta/2]$.
The model uses parallel updating, and,
in this paper, I will also assume the so-called standard Vicsek-model which uses a forward-upating rule.
This amounts to using the already updated positions ${\bf x}_i(t+\tau)$  
in determining the average directions $\Phi_i$.
The noise strength $\eta$ is an important parameter of the VM.
Another relevant parameter is the average particle number $M$ that can be found inside a circle of radius $r_0$. Thus, 
$M=\rho_0 \pi r_0^2$ where $\rho_0$ is the global number density. On the mean-field level, 
the average number of particles a particular focal particle is encountering inside
its collision circle,
is also given by $M$. 
The number density, $\rho_0$, itself is not a relevant parameter
because the particles have zero volume and the spatial ``extension'' of a particle is described by the range of the alignment interaction, $r_0$, 
instead.
One of the few meaningful dimensionless quantities to be formed by the number density is given by $M$, 
which describes the ratio of the interaction range to the average particle distance $1/\sqrt{\rho_0}$.

\section{Kinetic theory} 

In this paper the details of how to systematically derive kinetic and hydrodynamic equations for particle-based 
models with a discrete
time dynamics and very general collision rules are presented.
No linearizations or BGK-relaxation-type approximations (Bhatnagar-Gross-Krook), \cite{BGK_54}, are required.
Consequently, 
nonlinearities and gradient terms of the hydrodynamic fields can in principle be recovered to arbitrary order
with no free parameters. % \cite{FOOT1}. 
The main approximations involved are the assumption of Molecular chaos and that the system is not 
too inhomogeneous. The former assumption amounts to a mean-field approach; the latter means that gradients in 
the hydrodynamic variables are small enough to justify a gradient expansion.

The kinetic approach presented here relies on a properly designed ensemble of macroscopically identical copies 
of the system. 
This allows the definition of the N-particle probability density,
$P(\theta^{(N)}, {\bf X}^{(N)},t)$ 
where 
${\bf X}^{(N)} \equiv({\bf x}_1,{\bf x}_2,..., {\bf x}_N)$,
and
$\theta^{(N)} \equiv(\theta_1,\theta_2,..., \theta_N)$.
Here, the velocities
${\bf V}^{(N)} \equiv({\bf v}_1,{\bf v}_2,..., {\bf v}_N)$, are given in terms
of the angles, ${\bf v}_i=v_0(\cos{\theta_i}, \sin{\theta_i})$.
Then, $P_N\,\prod_i\,d{\bf x}_i\,d\theta_i$ gives the fraction of those members of the ensemble within which
particle $1$ is found in the phase space element $d{\bf x}_1\,d\theta_1$ around position ${\bf x}_1$ and angle $\theta_1$ while
simultaneously particle $2$ is in the element $d{\bf x}_2\,d\theta_2$ around position ${\bf x}_2$ and angle $\theta_2$, and so on.
Macroscopic quantities, such as the density field $\rho({\bf r},t)$ or the two-time, two-point correlation function
$g_2({\bf r},\phi,{\bf r}',\phi',t')$ can then be defined as ensemble averages 
over their microscopic counterparts.
To describe quantities that merely depend on one position and one time, only the one-point
microscopic phase space density
\begin{equation}
\label{PSI1}
\Psi_1=\sum_i\delta(\theta-\theta_i)\delta({\bf x}-{\bf x}_i)
\end{equation}
is needed. See Refs. \cite{klimon_67_74,ernst_81} for more information on phase space densities.
When viewed at fixed ${\bf x}$ and $\theta$ for one particular member of the ensemble,  
$\Psi_1$ is strongly fluctuating in time $t$ due to the underlying particle 
dynamics. 
The ensemble average over $\Psi_1$ defines the one-particle density $f$,
\begin{equation}
\label{FDEFIN}
f({\bf x},\theta,t)=\big\langle \Psi_1\big\rangle=
\int \Psi_1 \,P(\theta^{(N)}, {\bf X}^{(N)},t)\prod_i^N\,d{\bf x}_i\,d\theta_i\,,
\end{equation}
Additional integrations over the angle $\theta$
yield the macroscopic number and momentum density, $\rho({\bf x},t)$ and ${\bf w}({\bf x},t)$, respectively:
\begin{eqnarray}
\nonumber
\rho({\bf x},t)&=&\int_0^{2\pi} f({\bf x},\theta,t)\,d\theta \\
\label{FMOMENT}
{\bf w}({\bf x},t)&=&\int_0^{2\pi} {\bf v}\, f({\bf x},\theta,t)\,d\theta 
\end{eqnarray}
with ${\bf v}=v_0(\cos{\theta},\sin{\theta})$.
In that sense, 
$\rho({\bf x},t)$ denotes the average density at position ${\bf x}$, where the average is performed 
over all members of the ensemble. 
The goal of a mean-field kinetic theory is to first derive a closed equation for $f({\bf x},\theta,t)$, which
is then either analyzed directly or used to derive hydrodynamic equations for macroscopic fields such as $\rho$ and ${\bf w}$.

The starting point of the kinetic approach presented here, called phase space approach (PSA) in Ref. \cite{ihle_14_a}, 
is not  
an approximate equation for $f$ such as the Boltzmann-, Enskog- or Vlasov-equations. 
Instead, one starts at a more fundamental level with an exact
evolution equation for the N-particle probability density $P_N$, 
\begin{equation}
\label{LIOU_SWARM1}
P(\theta^{(N)}, {\bf X}^{(N)}+\tau {\bf V}^{(N)},t+\tau)=C\diamond 
P(\theta^{(N)}, {\bf X}^{(N)},t) 
\end{equation} where the collision operator $C$ encodes the microscopic
collision rules,
\begin{eqnarray}
\nonumber
C\diamond P(\theta^{(N)}, {\bf X}^{(N)},t)&=&
{1\over \eta^N}
\int_{-\eta/2}^{\eta/2}
d\xi^{(N)}
\int_0^{2\pi} d\tilde{\theta}^{(N)}
\,P(\tilde{\theta}^{(N)}, {\bf X}^{(N)},t)
\\
\label{COLL_INT_SWARM1}
&\times & \prod_{i=1}^N \hat{\delta}(\theta_i-\xi_i-\Phi_i(\tilde{\theta}^{(N)}, {\bf X}^{(N)};r_0))\,.
\end{eqnarray}
Here,
$\hat{\delta}(x)=\sum_{m=-\infty}^{\infty}\delta(x+2\pi m)$
is the periodically continued delta function which
is needed to map angles outside the range $[0,2\pi]$ back to this interval.
The mean angle $\Phi_i$ depends on the precollisional angles $\tilde{\theta}_i$ of the particles which are located
in a circle of radius $r_0$ around particle $i$. Thus, $\Phi_i$ depends also implicitly on the interaction range $r_0$
and on the positions of the particles. This mean angle $\Phi_i$ is defined in Eq.~(\ref{COLLIS}).
The first integral in Eq.~(\ref{COLL_INT_SWARM1}) averages
over the individual angular noises, $\xi_i$.

Eq.~(\ref{LIOU_SWARM1}) describes a Markov-chain in phase space.
It is exact and contains the microscopic details of the multi-body collision
rules introduced in Sec.~\ref{sec:Vicsek}.
An alternative but equivalent approach is to construct an exact 
evolution equation for the microscopic 
quantity $\Psi_1$ itself. This approach was pioneered by Klimontovich \cite{klimon_67_74} 
and is often used in plasma physics 
\cite{nicholson_83}. 
Very recently, it has also been applied to Vicsek-like models \cite{chepizhko_14}.
However, I believe that complicated multi-particle collisions which cannot be expressed as a sum of pairwise interactions are 
technically easier to treat in the current approach, even at mean-field level. 
This belief is supported by the observation that the authors of Ref. \cite{chepizhko_14}
chose to modify the interaction rules of the VM to make them pairwise additive, which is not required for PSA.

Because the phase space approach is based on an ensemble average over infinitely many members, 
the resulting kinetic and hydrodynamic equations do not and must not contain
explicit noise-terms. This is because whether noise terms are required or not depends on
the way how the main kinetic quantity, which is very often denoted by the same symbol $f({\bf x},\theta)$, is defined. 
If one were to derive an approximate equation for the strongly fluctuating 
microscopic density $\Psi_1$ itself, for example by defining $\tilde{f}=\langle\Psi_1\rangle+\mu$ where $\mu$ is supposed to model 
the fluctuating difference $\Psi_1-\langle\Psi_1\rangle$ which 
must be noisy by definition, the resulting kinetic equation for $\tilde{f}$ would need an explicit noise term. 
In this work, the final kinetic equation is for the one-particle density $f$ defined in Eq. (\ref{FDEFIN}), whose evolution
equation is noise-free. However, if going beyond the mean-field approximation \cite{chou_15} , 
this equation will also 
depend on two-point, three-point and higher 
correlation functions.
A thorough discussion on when noise terms are required can be found in Ref. \cite{archer_04}.

Similar to the BBGKY-hierarchy in classical mechanics, it is possible to derive a hierarchy
of evolution equations for $n$-particle correlation functions with $n=1,2\ldots N$.
from Eqs.~(\ref{LIOU_SWARM1},\ref{COLL_INT_SWARM1}).
In Ref.~\cite{chou_15} the first two members of this hierarchy were considered.

In this paper, I use the simplest and most common way to close this hierarchy and
assume that there are no correlations 
among particles {\it prior} to the collisions.
This means that the probability distribution $P$
just before the collision step is approximated by a product of identical one-particle probability distributions:
\begin{equation}
\label{PROD_ANSATZ}
P(\theta^{(N)}, {\bf X}^{(N)}, t)
=\prod_{i=1}^N P_1(\theta_i, {\bf x}_i,t)
\end{equation}
This is the assumption of
Molecular Chaos which is reasonable at moderate to large noise strength $\eta$ and/or at a large
ratio of mean free path, $\Lambda=v\,\tau$, to the radius of interaction, $r_0$.
More discussions on the validity of this Ansatz in active matter can be found in Refs. \cite{chou_15,thueroff_13,hanke_13}.
On the technical level, for a time-discrete model such as the VM, the factorization is only used on the r.h.s of 
Eq. (\ref{LIOU_SWARM1}).
The Molecular Chaos approximation has a long history in the kinetic theory of gases. 
Its proper use, \cite{FOOT1}, in a formal derivation of the Boltzmann equation for a regular gas with 
time-continuous dynamics from the BBGKY-hierarchy
is quite intricate and involves further assumptions, see for example
Refs. \cite{grad_58,kreuzer_81,cercignani_88}.

Using the factorization, Eq.~(\ref{PROD_ANSATZ}), an Enskog-like equation for the one-particle probability distribution
can be obtained.
This is achieved by first multiplying Eq. (\ref{LIOU_SWARM1}) by 
the phase space density $\Psi_1$, Eq.~(\ref{PSI1}), and subsequently integrating 
over all particle positions $x_i$ and angles $\theta_i$.
This amounts to an ensemble average of $\Psi_1$  and leads to the distribution function $f(\theta,{\bf x})$.
The set of variables $(\theta,{\bf x})$ will be called field variables which have to be distinguished from 
the phases of individual particles, $(\theta_i,{\bf x}_i)$.
Each term in the sum, Eq. (\ref{PSI1}), gives only a non-zero contribution if a particular particle $j$ 
resides at location ${\bf x}$, that is ${\bf x}_j={\bf x}$.
This particle is called the {\em focal} particle. Since all $N$ particles behave the same and have the same physical properties,
it suffices to merely consider the first term in $\Psi_1$. This amounts to taking particle 1 as the focal particle, that is, $j=1$ and
${\bf x}_1={\bf x}$. The contributions from the other terms in Eq. (\ref{PSI1}) are included by an overall factor of $N$.
This factor of N is incorporated into the definition of the one-particle distribution function,
$f(\theta,{\bf x},t)=NP_1(\theta,{\bf x},t)$, and, in the thermodynamic limit $N\rightarrow \infty$ will drop entirely from the kinetic equations.

Although the position and the post-collisional angle $\theta_1$ of particle 1 is now fixed in our calculation,
all possible configurations of the remaining $N-1$ particles have to be considered.
The post-collisional angle of the focal particle depends on these configurations.
For example, if only particle 3 is inside the collision circle of particle 1, the alignment rule describes a binary collision and
has a different effect
as if, for example, particles 2, 6, and 9 were within interaction range and participate in a four-particle interaction.
In order to properly sum over all possible configurations, the integration over the locations of particles $2,3,\ldots N$ is split 
into integrations over particles {\em inside} the collision circle around the focal particles, and integrals over positions that are {\em outside}.
Mathematically, this is described by the identity 
\begin{eqnarray}
\nonumber
& &\int_{\textrm{all\;\;space}} d{\bf x}_2\,
d{\bf x}_3\,
d{\bf x}_4
\ldots
d{\bf x}_N\rightarrow \\
& &
\sum_{n=1}^N\, C_n\,
\int_{\textrm{out}}
d{\bf x}_{n+1}\,
d{\bf x}_{n+2}\,\ldots
d{\bf x}_N\,
\int_{\textrm{in}}
d{\bf x}_2\,
d{\bf x}_3\,\ldots
d{\bf x}_n
\enspace 
\end{eqnarray}
where $C_n$ is a combinatorial factor. This factor counts the number of possibilities 
$n-1$ particles can be picked from the available particles $N-1$ particles and assigned to the collision circle,
yielding
\begin{equation}
C_n= {(N-1)! \over (n-1)! \,(N-n)!}\enspace .
\end{equation}
The subscripts ``$\textrm{in}$'' and ``$\textrm{out}$'' describe integrations over the inside and outside of the collision circle, respectively.

Consider a location ${\bf x}$ which will also be the location of the focal particle $j=1$.
Because the particles have zero volume, the collision circle around $\bf x$ 
can contain any particle number $n$ between one and N. There
are $C_n$ possibilities to assign n particles to that circle.
For every such microstate, one integrates over the positions 
and angles of all other $N-n$ particles outside this circle.
This is easily done, because the collisions that affect particle $1$ only couple the $n$ particles inside its interaction circle.
Assuming an inhomogeneous density distribution 
$\rho(x)$, Eq.~(\ref{FMOMENT}),
and integrating one particle over all pre-collisional angles and all possible positions {\em outside} the collision circle  
in the area $\tilde{A}=A-\pi r_0^2$
gives the contribution
\begin{eqnarray}
\nonumber
& &\int_0^{2\pi} \,d\theta\int_{\tilde{A}} d{\bf x}\, P_1({\bf x},\theta)=
{1\over N}\int_0^{2\pi} \,d\theta\int_{\tilde{A}} d{\bf x}\, f({\bf x},\theta)= \\
\label{DENS_FUNCTIONAL1}
& &
{1\over N}\int_{\tilde{A}} \,d{\bf x}\, \rho(x,t)
={1\over N}\left(N-\int_{\odot}\,dx\rho(x,t) \right)
\end{eqnarray}
where the subscript $\odot$ denotes an integration over the collision circle centered around ${\bf x}={\bf x}_1$.
These integrations over the $N-n$ particles outside the circle lead to a total factor $(1-M_R({\bf x})/N)^{N-n}$, 
where $M_R({\bf x})$ is the position-dependent average number of particles in the collision circle,
\begin{equation}
\label{DENS_FUNCTIONAL2}
M_R({\bf x},t)=\int_{\odot}\rho({\bf x}_2,t)\,d{\bf x}_2
\end{equation}
Note, 
that $M_R$ is actually a {\em functional} of the density, $\rho$, and hence also a functional of the distribution function $f$ itself. Thus, $M_R$ can differ from 
the constant global value $\langle M_R\rangle=M=N/A$.
Neglecting this seemingly small difference would lead to spurious
gradient terms in the macroscopic equations. 

In the large N limit, one obtains the Poisson distribution
for the number of particles in a circle, 
\begin{equation}
{N!\over n! (N-n)!} \left(1-{M_R\over N}\right)^{N-n}\approx {N^n\over n!} {\rm e}^{-M_R}\,,
\end{equation}
and finally arrives 
at an Enskog-like equation,
\begin{equation}
\label{ENSKOG_MAIN}
f({\bf x}+\tau {\bf v},\theta,t+\tau)=C_E \diamond
f({\bf x},\theta,t))\,, 
\end{equation}
where $C_E$ is an Enskog collision operator for multi-particle collisions defined by 
\begin{eqnarray}
\nonumber
C_E\diamond f({\bf x},\theta, t)&=&
{1\over \eta}
\int_{-\eta/2}^{\eta/2}
d\xi
\bigg\langle \bigg\langle
\sum_{n=1}^N
{{\rm e}^{-M_R}\over n!}
\,n\, \\
\label{ENSKOG1}
& &\times f(\tilde{\theta}_1,{\bf x}, t)
\,\hat{\delta}(\theta-\xi-\Phi_i)
\,\prod_{i=2}^n f( \tilde{\theta}_i, {\bf x}_i,t)
\bigg\rangle_{\tilde{\theta}} \bigg\rangle_x
\,.
\end{eqnarray}
Here,
$\langle ... \rangle_x=\int_{\odot}... \,d{\bf x}_2\,d{\bf x}_3...d{\bf x}_n$
denotes the integration over all positions of the particles $2,3,\ldots n$ inside the collision circle, 
and $\langle ... \rangle_{\tilde{\theta}}=\int_0^{2\pi}... d\tilde{\theta}_1\, d\tilde{\theta}_2\ldots d\tilde{\theta}_n$
refers to the integration over the pre-collisional angles of all $n$ particles inside the circle.

The form of the collision operator can be understood as follows:
(i) it is a sum over all possible particle numbers $n$ that could be found inside the interaction circle of particle $1$,
weighted by the Poisson-probability. The Poisson-property is a result of the Molecular chaos approximation.
(ii) the additional factor $n$ accounts for the fact that it can be any of the $n$ particles that is pinned
at position $x$ and whose direction is updated to $\theta$ in the collision, i.e. there is n possibilities to pick one of the 
particles,
(iii) one has to integrate over all the other $n-1$ particles's positions inside the circle and all 
possible pre-collisional velocities which would result in the desired outcome-direction $\theta$ of the focal particle,
(iv) finally there is an average over the distribution of the noise that is applied to the focal particle. 

The consistency of the kinetic equation can be tested by
setting all distribution functions in $C_E\diamond f$ equal to the homogeneous, disordered solution $f_0=\rho_0/(2\pi)$, 
with a homogeneous density $\rho_0=\int_0^{2\pi} d\theta f_0=\rho_0$. 
Then, all integrations and summations
in the collision integral can be performed
exactly
by using the series expression of the exponential function, $e^{M}=\sum_n M^n/n!$ and the integral representation of the $\delta$-function,
\begin{equation}
\label{DELTA_EXP}
\delta({\bf v}-{\bf v}_0)=\int_{-\infty}^{\infty}\, {\rm e}^{i{\bf k}({\bf v}-{\bf v}_0)}\, {d{\bf k}\over (2\pi)^d} 
\end{equation}
One finds that $f_0$ is a fixed point of the integral equation,
$f_0=C_E\diamond f_0$, thus $f_0$ is a homogeneous steady state distribution
of the VM at all values of the noise $\eta$.
Not passing this test would mean that the collision operator violates mass conservation.

\section{Deriving hydrodynamic equations}
\label{subsec:Hydro}

\subsection{Background}

Extracting macroscopic behavior
from kinetic equations has a long tradition in the kinetic theory of gases and 
plasmas.
One of the methods to achieve this is the Chapman-Enskog expansion (CE) which 
involves expansions
in small temporal and spatial gradients of hydrodynamic 
fields \cite{enskog_21,chapman_52,hirschfelder_54,mcquarrie_76}.
The key assumption is that after a few collisions which 
can involve rapid changes of the distribution function $f$, 
the system reaches a ``hydrodynamic state''
where local equilibrium is achieved. In this state, 
$f$ is assumed to be a functional of the slow hydrodynamic fields 
and should depend only indirectly on space and time through those fields. 
The hydrodynamic variables are the lowest angular moments of $f$, such as density and momentum density.
The Chapman-Enskog assumption is equivalent to the claim that 
the first few moments suffice to describe the system on large length and time scales. 

Since $f$ is uniquely defined by all its moments, this assumption would be justified if either all higher moments are 
negligibly small or that they are ``enslaved'' to the lower moments, meaning that they could be expressed as functionals
of the lower moments. 
Enslavement of the higher modes can be assured if there is a clear separation of time scales.
This is usually achieved by explicitly keeping those modes in the Chapman-Enskog expansion that either 
fulfill a conservation law or become soft
close to a transition. Hence, the choice of the first few moments is not arbitrary in CE.  

%These first relevant modes are given by conservation laws and soft modes close to a transition. 
%This leads to a separation
%of time scales between these slow modes and the higher fast modes.

CE takes
the local stationary state as a reference state and expands
around it in powers of the hydrodynamic gradients.
To systematically account for these gradients
a dimensionless ordering parameter $\epsilon$ is introduced, which is set to unity at the end
of the calculation. The physical meaning of this parameter is that it
is supposedly proportional to the Knudsen number, e.g.
the ratio of the mean free path to the length scale over which hydrodynamic fields change considerably.

\subsection{Chapman-Enskog expansion}
\label{subsec:Chapman}

The Chapman-Enskog
procedure
starts with 
a Taylor expansion of the l.h.s of Eq. (\ref{ENSKOG_MAIN})
around $({\bf x},\theta, t)$,
\begin{equation}
\label{ENSKOG_START}
f({\bf x}+\tau {\bf v},\theta,t+\tau)=\sum_{k=0}^{\infty}
{\tau^k(\partial_t+e_{\alpha}\partial_{\alpha})^k\over k!}\,
f({\bf x},\theta,t))\,,
\end{equation}
with ${\bf v}=(e_x,e_y)$.
Then, spatial gradients are scaled as
$\partial_{\alpha}\rightarrow \epsilon\partial_{\alpha}$, 
and
multiple time scales $t_i$ are introduced in the temporal gradients.
For the VM, the following scaling 
was chosen,
\begin{equation}
\label{TIME_SCALING}
\partial_t=\partial_{t_0}+\epsilon\partial_{t_1}+\epsilon^2\partial_{t_2}+\epsilon^3\partial_{t_3} \ldots\, .
\end{equation}
This choice 
differs from the usual set of equations for fluid flow \cite{frisch_87,chen_98,ihle_00,ihle_09}
because of the fast time scale $t_0$ which is not multiplied by a power of $\epsilon$ and
contributes time derivatives of all orders. 
The reason to introduce this time scale is that momentum is not locally conserved in the VM. 
Therefore,
the macroscopic momentum transport equation must have a source term; it cannot be written as a continuity equation, and
a new non-hydrodynamic
time scale should come into play. Momentum can still change even if spatial gradients of density and momentum
are zero. Therefore, in a gradient expansion with expansion parameter $\epsilon$, this new scale
must be of order $O(\epsilon^0)$ to be consistent with the assumption that 
spatial gradients are proportional to some non-zero power of $\epsilon $. 
Expansions that
contain all powers of $\partial_{t_0}$
can be conveniently summed up by the time evolution operator
\begin{equation}
\label{TIME_EVOLV0}
T\equiv{\rm exp}\left(\tau \partial_{t_0}\right)
\end{equation}
which shifts the time-argument of a function by the discrete time step $\tau$,
$T\circ f(t)=f(t+\tau)+O(\epsilon)$.

The CE proceeds with expanding the distribution function $f$ and the collision integral $C_E$, e.g. the right hand side 
of Eq. (\ref{ENSKOG_MAIN}), in powers 
of $\epsilon$, 
\begin{eqnarray}
\nonumber
f&=&f_0+\epsilon f_1 +\epsilon^2 f_2+\epsilon^3 f_3+\ldots \\ 
C_E&=&C_0+\epsilon C_1+\epsilon^2 C_2+\epsilon^3 C_3+\ldots\,. 
\end{eqnarray}
Inserting this into 
Eqs.\ (\ref{ENSKOG1}, \ref{ENSKOG_START}), and collecting terms of the same order
in $\epsilon$ up to third order yields a hierarchy of evolution equations for the $f_i$.
\begin{eqnarray}
\label{HIERARCHY_SWARM1}
O(\epsilon^0):\;\;& &Tf_0=C_0 \\
\label{HIERARCHY_SWARM1b}
O(\epsilon^1):\;\;& &T\circ\left[f_1+Lf_0\right]=C_1 \\
\label{HIERARCHY_SWARM1c}
O(\epsilon^2):\;\;& &T\circ\left[f_2+Lf_1+{1\over 2!}L^2f_0+\tau \partial_{t_2} f_0\right]=C_2 \\
\label{HIERARCHY_SWARM1d}
O(\epsilon^3):\;\;& &T\circ\left[f_3+Lf_2+{1\over 2!}L^2f_1+{1\over 3!}L^3f_0+
\tau\partial_{t_3}f_0+\tau\partial_{t_2}\circ Lf_0+\tau\partial_{t_2}f_1 \right]=C_3~~~~~~~~~~~\\
\nonumber
& &\;\;{\rm with}\;\; D\equiv\tau\left(\partial_{t_1}+e_{\alpha}\partial_{\alpha}\right)
\end{eqnarray}
All spatial derivatives are contained in the ``convective'' time derivative $D$ which is of order $\epsilon$.
Due to the absence of momentum conservation and Galilean invariance this
set 
of equations is very different from the usual one. 
It is not {\em a priori} evident whether the scaling ansatz 
for the time derivatives is
correct.
However, it turns out that this choice is compatible with the microscopic collision rules
and avoids inconsistencies 
if additionally 
the expansion of the distribution function $f$ is
identified as an angular Fourier series,
\begin{eqnarray}
\label{FOURIERA1}
f&=&f_0+\epsilon f_1+\epsilon^2 f_2+\epsilon^3 f_3 ...\;\;{\rm with} \\
\label{FOURIERB1}
f_0({\bf x},t)&=&{\rho({\bf x},t) \over 2\pi} \\
\label{FOURIERC1}
f_n({\bf x},\theta,t)&=&{1\over \pi v_0^n}\left[a_n({\bf x},t)\cos{(n\theta)}+b_n({\bf x},t)\sin{(n\theta)}\right]
\;\;{\rm for}\;n>0\,,
\end{eqnarray}
if the analysis is restricted to the vicinity of the order-disorder transition,
and if a redefinition of the collision integrals $C_1$ and $C_3$ is performed, as explained later in 
Sec.~\ref{sec:Closure}.  
Hence, the reference state $f_0$ of the CE, i.e. the leading order contribution to $f$, 
coincides
with the zero mode of the Fourier series. 
Thus, the Chapman-Enskog expansion is performed around the disordered state where
particles have 
no preferred
direction. 

We are seeking a hydrodynamic description of the first two moments of $f$, namely the particle density $\rho$
and the momentum density vector ${\bf w}$,
\begin{eqnarray}
\rho&=&\int_0^{2\pi}f\,d\theta=
\int_0^{2\pi}f_0\,d\theta \\
w_x&=&\rho u_x=\int_0^{2\pi}e_x f\,d\theta=
\int_0^{2\pi}v_0\cos{\theta} f_1\,d\theta=a_1 \\
w_y&=&\rho u_y=\int_0^{2\pi}e_y f\,d\theta=
\int_0^{2\pi}v_0\sin{\theta} f_1\,d\theta=b_1
\end{eqnarray}
where the  microscopic velocity
vector is given by 
\begin{equation}
{\bf v}\equiv (e_x,e_y)=v_0(\cos{\theta},\sin{\theta})\,.
\end{equation}
The particular choice to expand in a Fourier-series means that the density is given by the lowest
order distribution function $f_0$ alone, and the momentum density is described by the next higher order term, $f_1$.
This is different to usual hydrodynamic models like the Lattice-Boltzmann method for an simple fluid
\cite{chen_98,mcnamara_93}, where the momentum density is given by a moment of $f_0$ and not $f_1$.
Thus, in first order one finds,
\begin{equation}
\label{FIRST_ORDER}
f={\rho\over 2\pi}+{1\over \pi v_0}\left[ w_x\cos{\theta}+w_y\sin{\theta}\right] +O(\epsilon^2)
\end{equation}
Multiplying the hierarchy of evolution equations (\ref{HIERARCHY_SWARM1}--\ref{HIERARCHY_SWARM1d}) 
by powers of the microscopic velocity
vector ${\bf v}$
and integrating over $\theta$ gives a set of equations
for the time development of the density and the moments $a_i$ and $b_i$.
A number of moments of $f$ and the collision operator $C_E$ 
occur in these equations and will be evaluated in the next sections.
After this, the nontrivial closure of the hierarchy of moment equations will be discussed and equations
for the hydrodynamic fields will be derived.

\subsection{Moments of the distribution functions and the collision operator}
\label{sec:Moments}

The following moments are needed in the derivation of the macroscopic equations:
\begin{eqnarray}
\label{MOMENT1}
\langle e_{\alpha}e_{\beta}f_0\rangle&=&\delta_{\alpha\beta}{v_0^2\over 2} \rho \\
\label{MOMENT2}
\langle e_{\alpha}e_{\beta}f_2\rangle&=&(\delta_{\alpha\beta x}-\delta_{\alpha\beta y }){a_2\over 2}+
(1-\delta_{\alpha\beta}){b_2\over 2} \\
\label{MOMENT3}
\langle e_{\alpha}e_{\beta} e_{\gamma} f_1\rangle&=&
(\delta_{\alpha\beta}w_{\gamma}
+\delta_{\alpha\gamma}w_{\beta}
+\delta_{\gamma\beta}w_{\alpha}){ v_0^2\over 4}
\\
\label{MOMENT4}
\langle e_{\alpha}e_{\beta} e_{\gamma} e_{\delta} f_0\rangle&=&
(\delta_{\alpha\beta}\delta_{\gamma\delta}+
\delta_{\alpha\gamma}\delta_{\beta\delta}+
\delta_{\alpha\delta}\delta_{\beta\gamma})
{v_0^4 \rho\over 8}
\end{eqnarray}
The brackets denote angular integration, $\langle\ldots\rangle=\int_0^{2\pi}\ldots d\theta$. 
The $f_n's$ are given in Eq. (\ref{FOURIERC1}). They depend on the angular Fourier coefficients 
of $f$ but also depend
on the cosine and sine of the angle $n\theta$.
Therefore, odd moments of $f_0$ such as $\langle e_{\alpha} f_0 \rangle$ and even moments of $f_1$ such as
$\langle f_1 \rangle$ and $\langle e_{\alpha} e_{\beta} f_1 \rangle$ are equal to zero.

In addition to the moments of $f_n$,
the first few velocity moments of 
the collision operator are needed and are calculated in the next section.
In this paper I will restrict myself 
to the case of large mean free path; that is I assume that the mean travel distance $\Lambda=\tau v_0$
is large compared to the interaction radius $r_0$, \cite{FOOT2}.
Hence, the ``collisional contributions'' to the transport coefficients will be neglected here.
These contributions are known from the Enskog theory of dense fluids, see Ref. \cite{hanley_72}. 
They take into account the transfer of energy and momentum via the intermolecular potential and arise 
because the size of colliding molecules is not neglected anymore.
These collisional contributions have been studied in a variety of particle-based models, 
such as Multi-Particle Collision Dynamics \cite{gompper_08,ihle_05} and Direct Simulation Monte Carlo \cite{alexander_98} 
where they become important if the interaction range is larger than the mean free path.

The restriction to large $\Lambda$ is not a principal limitation of the current approach but will simplify 
the evaluations of the collision operator in the following section.

\vspace{0.3cm}

\subsubsection{First order moment: $\langle e_{\alpha} C_1\rangle$}

To evaluate moments of the collision operator, 
the Fourier series for $f$, Eq. (\ref{FOURIERC1}), is 
inserted into the collision operator, Eq.~(\ref{ENSKOG1}). Multiplication with powers of the velocities ${\bf v}=(e_x,e_y)$
and integration over the angle $\theta$ leads to the disappearance of
the Dirac delta function in the collision operator.
In the limit of $\Lambda \gg r_0 $, variations of $f$ across the interaction circle are neglected.
As a result, the integrations over the interaction
circle become trivial and just lead to powers of $A=\pi r_0^2$.
The remaining angular integrations lead to integrals such as $K_C^1$, which I call $K$-integrals.
These integrals are evaluated
in Sec.~\ref{sec:Integral_eval} and given in Table 1. 

Keeping only terms linear in $\epsilon$ leads to
\begin{eqnarray}
\label{REDRESS2}
\langle e_{\alpha} C_1\rangle&=&\lambda w_{\alpha}
\\
\label{LAMBDA_DEF}
\lambda&=&{4 \over \eta}{\rm sin}\left({\eta\over 2}\right)
{\rm e}^{-M_R} \sum_{n=1}^{\infty} {n^2 M_R^{n-1}\over n!}\,K_C^1(n)
\end{eqnarray}
$M_R$ is the 
mean local particle number, $M_R=\int_{\odot}\rho({\bf x})\,d{\bf x}$ where the integration goes
over the interaction circle.
The coefficient $\lambda$ can be simplified in the limit $M_R\rightarrow \infty$.
Recall 
the first moments of the Poisson distribution:
\begin{eqnarray}
\nonumber
\sum_{n=0}^{\infty}
{{\rm e}^{-M_R}\over n!}M_R^n&=1 \\
\label{POIS_SIMP}
\sum_{n=0}^{\infty}
{{\rm e}^{-M_R}\over n!}nM_R^n&=&
M_R
\end{eqnarray}
and express the sum in Eq. (\ref{LAMBDA_DEF}) as an average over the Poisson distribution
\begin{equation}
{\rm e}^{-M_R} \sum_{n=1}^{\infty} {n^2 M_R^{n-1}\over n!}\,K_C^1(n)
={\rm e}^{-M_R}\sum_{n=0}^{\infty} {M_R^n\over n!}\,h(n)
\end{equation}
by means of the function $h(n)=(n+1)\, K_C^1(n+1)$.
Expanding $h$ around $n=M_R$ gives
\begin{eqnarray}
\nonumber
{\rm e}^{-M_R}\sum_{n=0}^{\infty} {M_R^n\over n!}\,h(n)
&=&
{\rm e}^{-M_R}\sum_{n=0}^{\infty} {M_R^n\over n!}
(h(M_R) \\
\label{EXPAND_H}
&+&(n-M_R)h'(M_R)+O(h''(M_R))
\approx  h(M_R)\,.
\end{eqnarray}
because the term proportional to $h'$ vanishes due to Eq. (\ref{POIS_SIMP}).
Applying this idea to Eqs. ({\ref{LAMBDA_DEF}) and using the  
expression for $K_C^1(n)\sim \sqrt{\pi}/(4\sqrt{n})$ in  the limit $M_R\rightarrow \infty$, see Table 1,
gives the asymptotic expression,
\begin{equation}
\label{LAMBDA_DEF_ASYMP}
\lambda\sim{1\over \eta}{\rm sin}\left({\eta\over 2}\right)\,\sqrt{M_R \pi}
\;\;\;{\rm for}\;\;M_R\gg 1
\end{equation}
More details on the expansion of all transport coefficients in the large $M_R$ limit are given in Ref. 
\cite{ihle_15_a}.
In the opposite limit of low density, $M_R\ll 1$, one finds
\begin{equation}
\label{LAMBDA_DEF_SMALL}
\lambda\sim
{2\over \eta}{\rm sin}\left({\eta\over 2}\right)
\left(1+M_R\left[{4\over \pi}-1\right]+O(M_R^2) \right)
\end{equation}

\subsubsection{Second order moment: $\langle e_{\alpha} e_{\beta} C_2\rangle$}

First, let us evaluate the case $\alpha=\beta=x$.
The result is
\begin{eqnarray}
\nonumber               
\langle e_x^2 C_2\rangle&=&{1\over \eta} \sin{(\eta)}\sum_{n=0}^N
{{\rm e}^{-M_R}\over n!}n
\Big\{2n M_R^{n-1} K_{2c}^{11} a_2+4AM_R^{n-2} {n \choose 2}  \\
\label{SECOND_XX_START}
& & \times[K_{cc}^{11} w_x^2+K_{ss}^{11}w_y^2]
\Big\}
\end{eqnarray}
The angular integrals such as $K_{cc}^{11}$ and $K_{2c}^{11}$ are calculated in Sec.~\ref{sec:Integral_eval}.
Using $K_{cc}^{11}=K_{cs}^{12}$ and $K_{ss}^{11}=-K_{cs}^{12}$ one finds,
\begin{eqnarray}
\label{SECOND_XX_LAST}
\langle e_x^2 C_2\rangle&=&
{p\over 2} a_2+{q\over 2}(w_x^2-w_y^2) \\
\label{P_DEF1}
p&=&
{4\over \eta} \sin{(\eta)}\sum_{n=1}^N
{{\rm e}^{-M_R}\over n!}n^2
M_R^{n-1} K_{2c}^{11}(n) 
\\
\label{Q_DEF1}
q&=&
{4A\over \eta} \sin{(\eta)}\sum_{n=2}^N
{{\rm e}^{-M_R}\over n!}n^2(n-1)
M_R^{n-2} K_{cs}^{12}(n) 
\end{eqnarray}
with the auxiliary quantities $p$ and $q$.
According to Eq.~(\ref{P_INTERPRET}) 
in Sec.~\ref{sec:Green_Kubo}, the quantity $p$ can be physically interpreted as the decay rate
of the kinetic stress correlations.
By means of Eq. (\ref{EXPAND_H}) the sums in Eqs.~(\ref{P_DEF1}) and (\ref{Q_DEF1}) can be approximated for large $M_R$ as,
\begin{eqnarray}
\label{SECOND_AP_P}
p&\approx&
{4\over \eta} \sin{(\eta)}\,
M_R
K_{2c}^{11}(M_R) 
\\
\label{SECOND_AP_Q}
q&\approx&
{4A\over \eta} 
\sin{(\eta)}\,
(M_R-1)
 K_{cs}^{12}(M_R) 
\end{eqnarray}
Finally, since the asymptotic behavior for $K_{2c}^{11}$ and $K_{cs}^{12}$ is known, one finds
in the limit $M_R\rightarrow \infty$:
\begin{eqnarray}
\label{SECOND_FIN_P}
p&\sim& {1\over 2 \eta}\sin{(\eta)} \\
q&\sim& {A\over 2 \eta}\sin{(\eta)} 
\end{eqnarray}
Because the values of the $K$-integrals are also known for small $n=1,2,3,...$, see Sec.~\ref{sec:Integral_eval}, 
the behavior
at small density $M_R\ll 1$ can be extracted easily as,
\begin{eqnarray}
\label{SECOND_FIN_P}
p&\sim& {1\over  \eta}\sin{(\eta)}(1-M_R+O(M_R^2)) \\
q&\sim& {A\over  \eta}\sin{(\eta)}(1+M_R(-1+12 K_{cs}^{12}(3))+O(M_R^2))\\
 &=& 
{A\over  \eta}\sin{(\eta)}(1-0.41344\, M_R+O(M_R^2))
\end{eqnarray}
Note that in the low density limit, $M_R\ll 1$,  the main contribution to $p$ comes from the ``self-interaction''
of a single particle which just randomly changes its flying direction without 
interference from another particle.
In contrast, the first contribution to $q$
comes from a binary collision, i.e. from just having two particles in an interaction circle.
The ``self-interaction'', leading to the dominant term in Eq. (\ref{SECOND_FIN_P})
and also in Eq. (\ref{LAMBDA_DEF_SMALL}), might not be present
in other versions of the VM. For example, one could decide to not perform a ``collision''
if no other particle is found in a circle around a given particle.
For larger densities, this little algorithmic detail would not make a difference, but for low densities,
the coefficients $p$ and $\lambda$ would differ from Eq.(\ref{SECOND_FIN_P}) and (\ref{LAMBDA_DEF_SMALL}), 
respectively.
In fact, in this case, both $p$ and $\lambda$ would go to unity
for $M_R\rightarrow 0$.
The result $\lambda\rightarrow 1$ makes  perfect physical sense: 
On one hand, for $M_R\rightarrow 0$ we can assume that a given particle will never meet another one, thus
it would never change direction according to this modified collision rule. 
Hence, its momentum is exactly conserved.
On the other hand, the 
signature of exact momentum conservation is $\lambda=1$ which is what we found here.

Because of $\langle C_2\rangle=0$ and $e_x^2+e_y^2=v_0^2$ one sees that 
\begin{equation}
\langle e_y^2 C_2\rangle=-\langle e_x^2 C_2\rangle
\end{equation}

A calculation similar to the one for $\langle e_x^2 C_2\rangle$ but now with $\alpha=x$ and $\beta=y$
yields,
\begin{eqnarray}
\nonumber                
& &\langle e_x e_y C_2\rangle= \\
\label{SECOND_XY_START}
& &{1\over \eta} \sin{(\eta)}\sum_{n=0}^N
{{\rm e}^{-M_R}\over n!}n
\left\{2n M_R^{n-1} K_{2s}^{12} b_2+8AM_R^{n-2} {n \choose 2} K_{cs}^{12} w_x w_y
\right\}
\end{eqnarray}
Since $K_{2s}^{12}=K_{2c}^{11}$, see Sec.~\ref{sec:Integral_eval}, one finds,
\begin{equation}
\label{SECOND_XY_LAST}
\langle e_x e_y C_2\rangle=
{p\over 2} b_2+qw_x w_y
\end{equation}
where the coefficients $q$ and $p$ are identical to the ones calculated above.

\subsubsection{Third order moment: $\langle e_{\alpha} C_3\rangle$}

The expansion of the distribution function $f$ in Fourier modes, see Eq. (\ref{FOURIERC1}), is inserted in the collision operator, Eq.~(\ref{ENSKOG1}),
and only terms of order $\epsilon^3$ are kept.
As before,
in the limit of $\lambda \gg r_0$, variations of $f$ across the interaction circle are neglected.
Hence, 
all arguments ${\bf x}_i$ in $f$ can be formally replaced by ${\bf x}$, and the integrations over the interaction
circle become trivial.
The result is for $\alpha=x$:
\begin{eqnarray}
\nonumber
\langle e_x C_3\rangle&=&
{2\over \eta v_0^2 }\sin{\left({\eta\over 2}\right)}
\sum_{n=0}^N
{{\rm e}^{-M_R}\over n!}n
\Bigg\{
8A^2 M_R^{n-3} {n \choose 3} \\
\nonumber
& & \left. \times \left[ w_x^3 K_{ccc}^1+3 w_x w_y^2 K_{css}^1 \right]
\right.
\\
& & +4A M_R^{n-2} n(n-1) K_{c2c}^1
\left[ w_x a_2+w_y b_2\right]
\Bigg\}
\end{eqnarray}
With $K_{css}^1=K_{ccc}^1/3$ one finds
\begin{eqnarray}
\label{THIRD_C3_1}
\langle e_x C_3\rangle&=&
\Gamma w_x w^2+S(w_x a_2+w_y b_2) \\
\label{THIRD_C3_2}
\Gamma&=&
{8A^2\over 3 \eta v_0^2 }\sin{\eta\over 2}
\sum_{n=3}^N
{{\rm e}^{-M_R}\over n!}
n^2 (n-1)(n-2)M_R^{n-3} K_{ccc}^1(n)
\\
\label{THIRD_C3_3}
S&=&
{8A\over \eta v_0^2 }\sin{\eta\over 2}
\sum_{n=2}^N
{{\rm e}^{-M_R}\over n!}
n^2(n-1) M_R^{n-2} K_{c2c}^1(n) 
\end{eqnarray}
with the auxiliary quantities $\Gamma$ and $S$.
$A$ is the area of the interaction circle, $A=\pi r_0^2$.

The coefficients $\Gamma$ and $S$ can be simplified in the limit $M_R\rightarrow \infty$
by means of Eq. (\ref{EXPAND_H}).
\begin{eqnarray}
\label{THIRD_AP_2}
\Gamma&\approx&
{8A^2\over 3 \eta v_0^2 }\sin{\eta\over 2}
{(M_R-1)(M_R-2)\over M_R}  K_{ccc}^1(M_R)
\\
\label{THIRD_AP_3}
S&\approx&
{8A\over \eta v_0^2 }\sin{\eta\over 2}
(M_R-1) K_{c2c}^1(M_R) 
\end{eqnarray}
Finally, since the asymptotic behavior for $K_{ccc}^1$ and $K_{c2c}^1$ is known, see Table 1, one has
in the limit of large $M_R$:
\begin{eqnarray}
\label{THIRD_FIN_2}
\Gamma &\approx&
-{A^2\over 4 \eta v_0^2 }
\sqrt{\pi\over M_R}
\sin{\eta\over 2}
\\
\label{THIRD_FIN_3}
S&\approx&
-{A\over 4 \eta v_0^2 }
\sqrt{\pi\over M_R}
\sin{\eta\over 2}
\end{eqnarray}
As before, a low density expansion of the coefficients can also be done for $M_R\ll 1$,
\begin{eqnarray}
\label{THIRD_LOWM_R}
\Gamma &=&
{8A^2\over \eta v_0^2 }\sin{\eta\over 2} 
K_{ccc}^1(3)
\left(1+M_R[-1+4K_{ccc}^1(4)/3K_{ccc}^1(3)]+O(M_R^2)\right)\\
&\approx&
-0.55736 {A^2\over \eta v_0^2 }\sin{\eta\over 2} 
(1-0.3943\,M_R+O(M_R^2))\\
\label{THIRD_LOWM_S}
S&=&
{16A\over \eta v_0^2 }\sin{\eta\over 2}
K_{c2c}^1(2)
\left(1+M_R[-1+3K_{c2c}^1(3)/2K_{c2c}^1(2)]+O(M_R^2)\right)\\
&\approx&
-{8A\over 3\pi \eta v_0^2 }\sin{\eta\over 2}
\left(1-0.63893\,M_R+O(M_R^2)\right)
\end{eqnarray}

\subsection{Reevaluation and closure of the moment equations}
\label{sec:Closure}

Before actually deriving hydrodynamic equations, 
several technical issues need to be resolved.
Most Chapman-Enskog expansions are performed to second order and not to third order as done here.
This leads to a complication which is hardly discussed in the literature, probably because it only occurs at 
orders higher than two.
It is the fact that the time evolution operators in the Chapman-Enskog expansion do not typically commute 
(see Eq. (\ref{FINAL_RHO_EX}) as an example).
Writing the time derivative as a sum of different time derivatives
in Eq. (\ref{TIME_SCALING})
is equivalent to an splitting of the evolution equation into separate parts
 which have to be put together after the derivation has been completed. 
This artificial splitting seems to be the reason for the unusual noncommutative property of the time derivatives. 
In order to include noncommutativity but at the same time keeping the notation compact,
I use the ``$\circ$'' symbol whenever two time derivatives that need special consideration ``hit each other''. 
The $\circ$ symbol is a symmetrization operator and means that all time derivatives
in the term under consideration must be symmetrized.
For example, with this notation one would have, 
\begin{equation}
\partial_{t_0}^2\circ\partial_{t_1}=
\partial_{t_1}\circ \partial_{t_0}^2=
\partial_{t_0}\circ\partial_{t_1}\circ \partial_{t_0}=
{1\over 3}\left( 
\partial_{t_0}^2\partial_{t_1}+
\partial_{t_0}\partial_{t_1}\partial_{t_0}
+\partial_{t_1} \partial_{t_0}^2
\right)
\end{equation}
Three terms occur because there is three distinct permutations.
As another example, the expression $\partial_{t_0}^2\circ\partial_{t_1}\circ\partial_{t_2}$ 
stands for $4!/2!=12$ terms obtained by permuting the four operators 
$(\partial_{t_0}\,,\partial_{t_0}\,,\partial_{t_1}\,,\partial_{t_2})$.
Note, that the symmetrization is not forced but occurs naturally when carefully performing the Chapman-Enskog expansion in terms of
noncommutative time evolution operators.

The other technical subtlety is, that for reasons explained further below, $C_1$ and $C_3$, in Eqs. 
(\ref{HIERARCHY_SWARM1b}, \ref{HIERARCHY_SWARM1d}) are replaced by $\tilde{C}_1$ and $\tilde{C}_3$, 
\begin{eqnarray}
\nonumber
\tilde{C}_1&=&C_1-C_1\left(1-{1\over \lambda}\right) \\
\label{REDRESS1}
\tilde{C}_3&=&C_3+C_1\left(1-{1\over \lambda}\right) 
\end{eqnarray}
Eq.~(\ref{REDRESS2}) shows that the first moment of the ``undressed'' first order collision contribution $C_1$, 
is proportional 
to the momentum density,   
$\langle e_{\alpha} C_1 \rangle=\lambda w_{\alpha}$.
Analyzing the final macroscopic equations reveal the physical interpretation of the prefactor $\lambda$: It is 
an (ensemble-averaged) amplification factor of the  momentum density, where
$\lambda$ equal to unity means that the collisions conserve momentum. This is typically not the case
in the VM, unless directly at the order-disorder transition point.

The condition $\lambda=1$ is identical to the condition for the mean-field bifurcation of a homogeneous
disordered solution into an ordered solution \cite{ihle_11,ihle_14_a,chou_12}. 
For a given fixed noise $\eta$, one can find a critical mean particle number $M_{R,crit}$ by the condition 
$\lambda(\eta,M_{R,crit})=1$. This means that the local order parameter $w$ (which happens to be equal to
the macroscopic momentum) will grow from an infinitesimal  initial value if the local density is quenched above the critical density.
This growth is initially exponentially but will be saturated by nonlinear effects.
In the opposite scenario, $w$ will decay to zero if a homogeneous ordered system 
is quenched to a lower density $M_R<M_{R,crit}$.
In a first attempt to derive hydrodynamic equations without the redefinition of Eq. (\ref{REDRESS1}) 
it turned out that closing the hierarchical set of moment equations 
is especially easy near the order-disorder transition where $\lambda=1$. 
In the vicinity of this bifurcation, higher order moments of $f$ are small; they are enslaved
to lower order moments,
and a useful macroscopic  description is already obtained by the equation for the density (continuity equation) and 
the momentum density (Navier-Stokes-like equation).
Here, enslavement means that a given higher order moment such as $a_2$ is approximately given by
a function of lower order moments and their spatial derivatives, and that the difference between
the time derivative of this function and the true time derivative of the higher moment is of order $\epsilon$ or smaller.

Near the order-disorder phase transition, one has $1-\lambda\ll 1$, which can be expressed as
$1-\lambda=a_0+a_1\epsilon+a_2\epsilon^2+...$.
One has to keep in mind that $\epsilon$ is a formal {\it ordering parameter} which is set to unity
at the end of the calculation.
This leaves ambiguities in choosing the $a_i$, for example one could decide to use $a_0=a_1=0$ or
$a_0=a_2=0$.
Here, I choose $a_0=a_1=0$, which means $1-\lambda=O(\epsilon^2)$.
This choice can be justified {\em a posteriori} by 
analyzing the final equation for the momentum density, Eq. (\ref{CONSTDENSITY2}),
in a homogeneous stationary state, which gives $|\vec{w}|=\sqrt{(\lambda-1)/q_3}=O(\epsilon)$. 
Hence, in the current scaling, the momentum density is predicted to be of order $\epsilon$, 
which is consistent with Eq. (\ref{FIRST_ORDER}) and the identification of $f_n$ with the angular Fourier coefficients,
Eqs. (\ref{FOURIERA1},\ref{FOURIERC1}).
Other choices for $1-\lambda$ should in principle 
lead
to the same final
set of equations, 
however certain terms would show up at different orders of the derivation and the simple Fourier Ansatz for $f_n$ probably
would 
have to be modified.

Using the bare value $C_1$ in the first order expression Eq. (\ref{HIERARCHY_SWARM1b}) together with the assumption
$1-\lambda=O(\epsilon^2)$  would introduce an 
inconsistency since $C_1$ contains higher order contributions. 
Eq. (\ref{REDRESS1}) shifts this higher order part into $C_3$.
Note, that this is just a redistribution between different approximation levels since
the sum stays the same: $\tilde{C}_1+\tilde{C}_3=C_1+C_3$. 
The following derivations will assume proximity to the transition point.
Thus, the macroscopic equations derived in this work
are unlikely to be valid at very small noise -- far away from the transition.
According to Eqs. (\ref{REDRESS1}, \ref{REDRESS2}) one has now
\begin{equation}
\label{REDRESS3}
\langle e_{\alpha} \tilde{C}_1 \rangle= w_{\alpha}
\end{equation}
That is, at first order in $\epsilon$, the collisions conserve momentum in the ensemble-averaged sense, 
which is to be expected near the transition.

\subsection{Density and momentum evolution for time scales $t_0$ and $t_1$}

In this paragraph, the first two moments of the first few equations of the hierarchy, 
(\ref{HIERARCHY_SWARM1}-\ref{HIERARCHY_SWARM1c}),
are evaluated. The resulting equations for the time evolution of density $\rho$ and momentum density ${\bf w}$  will be used
later to simplify the higher order evolution equations.

Because of particle number conservation the collision operator has the following properties:
$\langle C_0\rangle=\rho$ and $\langle C_n\rangle=0$ for $n>0$. Here, the brackets denote an average 
over the angle
$\theta$, $\langle ...\rangle=\int_0^{2\pi}\ldots \,d\theta $.
Hence, if the zeroth moment, i.e. the angular average, of the set of equations (\ref{HIERARCHY_SWARM1}) is 
taken, one finds
\begin{eqnarray}
\label{HIERARCHY_SWARM2}
O(\epsilon^0):\quad& &T\rho=(1+\tau\partial_{t_0}+{\tau^2\over 2}\partial^2_{t_0}+...)\rho=\rho \\
\nonumber
O(\epsilon^1):\quad& &\langle T\circ Lf_0\rangle=
\\
\label{HIERARCHY_SWARM2b}
& & \tau\left\{ \partial_{t_1}\rho+{1\over 2}\left[\partial_{t_1}(\partial_{t_0} \rho)
+\partial_{t_0}(\partial_{t_1}\rho)\right]
\right. \\
& &+\left. {1\over 6}
\left[
 \partial_{t_0}^2(\partial_{t_1}\rho)
+\partial_{t_1}\partial_{t_0} (\partial_{t_0}\rho)
+\partial_{t_0}\partial_{t_1} (\partial_{t_0}\rho)
\right]+...
 \right\}=0 
\end{eqnarray}
The first equation is solved by 
\begin{equation}
\label{RHO_0}
\partial_{t_0}\rho=0
\end{equation}
Using this result, the solution of the second equation is
\begin{equation}
\label{RHO_1}
\partial_{t_1}\rho=0
\end{equation}
Next, the evolution equation for the momentum density ${\bf w}$ is derived by multiplying Eqs.~ 
(\ref{HIERARCHY_SWARM1}--\ref{HIERARCHY_SWARM1d}) by the velocity component $e_{\alpha}$
and integrating over $\theta$. In order $\epsilon^0$, the trivial, but consistent result $0=0$ occurs, because
the first moment of $C_0$ vanishes, $\langle e_{\alpha} C_0\rangle=0$, which can be proven easily.
In order $\epsilon^1$ one finds
\begin{equation}
T\left[w_{\alpha}+\tau\partial_{\beta}\langle e_{\alpha}e_{\beta}f_0\rangle\right]=\langle
 e_{\alpha} \tilde{C}_1 \rangle=w_{\alpha}
\end{equation}
Here, Eq. (\ref{REDRESS3}) was used, which states that momentum conservation is approximately
realized near the order-disorder transition; deviations from this conservation law will show 
up in higher order equations.
According to Eq. (\ref{MOMENT1}), $ \langle e_{\alpha}e_{\beta}f_0\rangle=\delta_{\alpha\beta}v_0^2 \rho_0/2$,
and Eqs. (\ref{TIME_EVOLV0},\ref{RHO_0}) it follows that
\begin{equation}
\label{EQ139}
{\rm exp}(\tau\partial_{t_0})w_{\alpha}+{\tau v_0^2\over 2}\partial_{\alpha}\rho=w_{\alpha}
\end{equation}
Because $\partial_{t_0}\rho=0$, this equation is solved by:
\begin{equation}
\label{EULER0}
\partial_{t_0} w_{\alpha}=-\partial_{\alpha}\left({v_0^2\over 2}\rho\right)
\end{equation}
which is the Euler-equation with an ideal gas pressure $p_{id}=k_B T\rho$ and temperature $k_B T=v_0^2/2$ (the particle mass is assumed to be one).
This is the expected result for two-dimensional particles with constant speed $v_0$ and kinetic 
energy $v_0^2/2$.
Note, that without the assumption of $1-\lambda\ll 1$, a momentum source term $\sim w_{\alpha}$ would appear,
\begin{equation}
\partial_{t_0} w_{\alpha}=\mu w_{\alpha}+\nu \partial_{\alpha}\rho
\end{equation}
and an interesting logarithmic dependence on $\lambda$ would follow,
\begin{eqnarray}
\mu&=&{1\over\tau}{\rm ln}\lambda \\
\nu&=&{v_0^2 \over 2}{{\rm ln}\lambda\over 1-\lambda}
\,,
\end{eqnarray}
which, however, is not used in the remainder of this paper.
Finally, an equation for 
$\partial_{t_1} w_{\alpha}$,
is obtained by taking the first moment of the $O(\epsilon^2)$ equation 
(\ref{HIERARCHY_SWARM1c}). Remembering that both $\partial_{t_0}\rho$ and
$\partial_{t_1}\rho$ vanish, one finds
\begin{equation}
\label{WT1}
\partial_{t_1} w_{\alpha}=0\,.
\end{equation}

\subsection{Density evolution for time scales $t_2$ and $t_3$}

Taking the zeroth moment of the $O(\epsilon^2)$ equation of the hierarchy gives
\begin{equation}
\label{RHO2_START}
O(\epsilon^2):\quad T\circ\left[\tau\partial_{\alpha}w_{\alpha}+
{1\over 2}\tau^2\partial_{\alpha}\partial_{\beta}\langle e_{\alpha}e_{\beta}f_0\rangle+\tau\partial_{t_2}\rho
\right]=0
\end{equation}
Inserting the second moment of $f_0$ given in Eq. (\ref{MOMENT1}) and using $\partial_{t_0}\rho=0$
results in
\begin{equation}
\label{RHO2_INT}
T\partial_{\alpha} w_{\alpha}
+
R \partial_{t_2}\rho+
{\tau v_0^2\over 4}\partial_{\alpha}^2\rho=0
\end{equation}
where a new time evolution operator $R$ was introduced,
\begin{equation}
\label{R_OP_DEFINE}
R=\left(1+{\tau\over 2!}\partial_{t_0}+{\tau^2\over 3!}\partial_{t_0}^2+...\right)
\end{equation}
The first term in Eq. (\ref{RHO2_INT}) can be replaced by applying the $T$-operator to Eq. (\ref{EULER0}), which gives
\begin{equation}
\label{EULER1}
T\partial_{\alpha}w_{\alpha}=\partial_{\alpha}w_{\alpha}-{\tau v_0^2\over 2}\partial_{\alpha}^2\rho\,,
\end{equation}
and insertion into Eq. (\ref{RHO2_INT}) yields, 
\begin{equation}
\label{RHO_INTE1}
\partial_{\alpha} w_{\alpha}
+
R\partial_{t_2}\rho-
{\tau v_0^2\over 4}\partial_{\alpha}^2\rho=0
\end{equation}
Finally, using $\partial_{t_0}\rho=0$ and $\partial_{t_0}w_{\alpha}=-v_o^2\partial_{\alpha}\rho/2$ this equation is solved 
by the expected continuity equation,
\begin{equation}
\label{RHO_2}
\partial_{t_2}\rho+\partial_{\alpha} w_{\alpha}=0
\end{equation}
Note, that all unphysical diffusive terms, $\sim \partial_{\alpha}^2\rho$, present in the starting 
expression, Eq. (\ref{RHO2_INT})
have exactly cancelled which is non-trivial in systems with forces and/or absence of momentum conservation.
A useful relation which follows from eq. (\ref{RHO_2}) or 
(\ref{RHO_INTE1}) is
\begin{equation}
\label{DEF_HELP1}
T\circ\partial_{t_2}\rho=R\partial_{t_2}\rho=-\partial_{\beta}w_{\beta}+{\tau v_0^2\over 4}\partial_{\beta}^2\rho
\end{equation}
In order to get the final density evolution equation, 
Eq. (\ref{WT1}) is needed to evaluate
the zeroth moment of the $O(\epsilon^3)$ equations.
After a short calculation one finds,  
\begin{equation}
\label{RHO_3}
\partial_{t_3}\rho=0
\end{equation}
Collecting all four different time derivatives, Eqs. (\ref{RHO_0},\ref{RHO_1},\ref{RHO_2},\ref{RHO_3}),
 the final result
for the density evolution is
\begin{equation}
\label{FINAL_RHO}
\partial_t\rho=
(\partial_{t_0}+
\partial_{t_1}+
\partial_{t_2}+
\partial_{t_3})\rho+O(\epsilon^4)=
-\partial_{\alpha} w_{\alpha}
+O(\epsilon^4)\,.
\end{equation}
We are now in position to discuss the noncommutativity of the time derivatives again.
A simple example is the commutator $[\partial_{t_0},\partial_{t_2}]$: when applied to the density $\rho$,
a non-zero result is obtained,
\begin{equation}
\label{FINAL_RHO_EX}
[\partial_{t_0}, \partial_{t_2}]\rho=
\partial_{t_0}\partial_{t_2}\rho-\partial_{t_2}\partial_{t_0}\rho=
{v_0^2\over 2}\nabla^2\rho
\,.
\end{equation}

\subsection{Momentum density evolution for time scale $t_2$}

To find $\partial_{t_2}w_{\alpha}$, Eq. (\ref{HIERARCHY_SWARM1d}) is multiplied by 
$e_{\alpha}$ and the angular average is taken.
This leads to
\begin{eqnarray}
\nonumber
& & T\left[
\partial_{\beta}\langle e_{\alpha} e_{\beta} f_2 \rangle+{\tau\over 2}\partial_{\beta}
\partial_{\delta}\langle e_{\alpha} e_{\beta} e_{\delta} f_1 \rangle
+{\tau^2\over 6}\partial_{\beta}\partial_{\delta}\partial_{\gamma}
\langle e_{\alpha} e_{\beta} e_{\gamma} e_{\delta} f_0 \rangle\right] \\
\nonumber
& + & T\circ\left[
{\tau\over 2}\partial^2_{t_1}w_{\alpha}+{\tau^2\over 2}\partial^2_{t_1}\partial_{\alpha}
\left({v_0^2\over 2}\rho\right)
+\tau\partial_{t_2}\partial_{\alpha}\left({v_0^2\over 2}\rho\right)
+\partial_{t_2}w_{\alpha}
\right] \\
\label{SECOND_W0}
& = &{\langle e_{\alpha} \tilde{C}_3\rangle\over \tau}
\end{eqnarray}
Using Eqs. (\ref{RHO_0}, \ref{RHO_1}, \ref{EULER0}, \ref{WT1}) one can show that the fourth and 
fifth term on the l.h.s.
vanish.
The sixth term is evaluated by means of Eqs. (\ref{DEF_HELP1}) as
\begin{equation}
\label{SECOND_SIX}
{\tau v_0^2\over 2}\partial_{\alpha}\left[ T\circ \partial_{t_2}\rho\right]=
{\tau v_0^2\over 2}\partial_{\alpha}
\left[ -\partial_{\beta}w_{\beta}+{\tau v_0^2\over 4}\partial^2_{\beta}\rho
\right]
\end{equation}
The evaluation of the seventh terms gives 
\begin{equation}
\label{SECOND_SEVEN}
T\circ \partial_{t_2}w_{\alpha}=R\,\partial_{t_2}w_{\alpha}+{\tau v_0^2\over 4}\partial_{\alpha}
\partial_{\beta}w_{\beta}-{\tau^2 v_0^4\over 24}\partial_{\alpha}\partial_{\beta}^2\rho
\end{equation}
where $R$ is defined in Eq. (\ref{R_OP_DEFINE}).
The moments in the first three terms are given in Eqs. (\ref{HIERARCHY_SWARM1b} -- \ref{HIERARCHY_SWARM1d}), and after inserting
everything, Eq. ({\ref{SECOND_W0}) for $\alpha=x$ becomes:
\begin{eqnarray}
\nonumber
& &T\left[
{1\over 2}(\partial_x a_2+\partial_y b_2)+{\tau v_0^2\over 8}(\nabla^2w_x+2\partial_x{\bf \nabla 
w})+{\tau^2v_0^4\over 16}\partial_x \nabla^2\rho\right] \\
\nonumber
&+&R\,\partial_{t_2}w_x+{\tau v_0^2\over 4}\partial_x{\bf \nabla w}
-{\tau^2v_0^4\over 24}\partial_x \nabla^2 \rho
+{\tau v_0^2\over 2}\partial_x\left\{-{\bf \nabla w}+{\tau v_0^2\over 4}\nabla^2\rho \right\} \\
\label{SECOND_SEVEN1}
&=&{\langle e_x\tilde{C}_3\rangle\over \tau}
\end{eqnarray}
Using $\partial_{t_0}\rho=0$ and $\partial_{t_0}w_x=-v_0^2\partial_x\rho/2$, one has
\begin{eqnarray}
\nonumber
T\partial_x\nabla^2\rho&=&\partial_x\nabla^2\rho \\
\nonumber
T\nabla^2w_x&=&\nabla^2w_x-{\tau v_0^2\over 2}\partial_x\nabla^2\rho\\
\label{SECOND_SIMPLI}
T\partial_x{\bf \nabla w}&=&\partial_x{\bf \nabla w}-{\tau v_0^2\over 2}\partial_x\nabla^2\rho
\end{eqnarray}
which helps simplifying the terms in brackets in Eq. (\ref{SECOND_SEVEN1})
with the result
\begin{equation}
\label{FINAL_SECOND_WX}
T\left[
{1\over 2}(\partial_x a_2+\partial_y b_2)\right]
+{\tau v_0^2\over 8}\nabla^2w_x
-{\tau^2v_0^4\over 24}\partial_x \nabla^2\rho
+R\,\partial_{t_2}w_x
={\langle e_x\tilde{C}_3\rangle\over \tau}
\end{equation}
In order to simplify the following analysis, a rescaling of time and space is used from now on.
Dimensionless spatial coordinates and dimensionless time are defined by
\begin{eqnarray}
\tilde{x}_{\alpha}&=&{x_{\alpha}\over \Lambda} \\
\tilde{t}&=&{t\over \tau}
\end{eqnarray}
where $\Lambda=v_0\tau$ is the mean free path. The density and momentum density are 
nondimensionalized accordingly by
\begin{eqnarray}
\nonumber
\tilde{\rho}_{\alpha}&=&\rho \Lambda^2 \\
\label{NONDIM1}
\tilde{w}_{\alpha}&=&w_{\alpha}{\Lambda^2\over v_0}
\end{eqnarray}
The operators $T$ and $R$, and the amplitudes $a_2$ and $b_2$ are rescaled consistently, 
\begin{eqnarray}
\nonumber
\tilde{T}&=&1+\partial_{\tilde{t}_0}+
\nonumber
{1\over 2 !} \partial^2_{\tilde{t}_0}+ 
\nonumber
{1\over 3 !} \partial^3_{\tilde{t}_0}+ ...\\
\nonumber
\tilde{R}&=&1+{1\over 2 !}\partial_{\tilde{t}_0}+
{1\over 3 !} \partial^2_{\tilde{t}_0}+
{1\over 4 !} \partial^3_{\tilde{t}_0}+...\\
\nonumber
\tilde{a}_2&=&a_2 \tau^2\\
\label{NONDIM2}
\tilde{b}_2&=&b_2 \tau^2
\end{eqnarray}
The coefficients occurring in the evaluation of moments are rescaled as 
\begin{eqnarray}
\nonumber
\tilde{\Gamma}&=&\Gamma {v_0^2\over \Lambda^4}\\
\nonumber
\tilde{S}&=&S {v_0^2\over \Lambda^2} \\
\label{NONDIM3}
\tilde{q}&=&{q \over \Lambda^2} 
\end{eqnarray}
The coefficients $\lambda$ and $p$ are already dimensionless.
To simplify notation, the tilde in the rescaled quantities will be omitted
in the remainder of the paper.

Using Eq. (\ref{THIRD_C3_1}) and rescaling, Eq. (\ref{FINAL_SECOND_WX}) now reads
\begin{eqnarray}
\nonumber
& &{T\over 2}(\partial_x a_2+\partial_y b_2)
+{1\over 8}\nabla^2w_x
-{1\over 24}\partial_x \nabla^2\rho
+R\,\partial_{t_2}w_x
\\
\label{FINAL_SECOND_WX_SCAL}
& &=\Gamma w_x w^2+S(w_x a_2+w_y b_2)
+(\lambda-1)w_x
\end{eqnarray}
The last term comes from the relation 
$\langle e_{\alpha}\tilde{C}_3\rangle
=\langle e_{\alpha}C_3\rangle+
\langle e_{\alpha}C_1\rangle (1-1/\lambda)$, see Eq. (\ref{REDRESS1}).
It is interesting to note that all terms containing the divergence of ${\bf w}$, 
$\sim \partial_x{\bf \nabla w}$ have cancelled like in the continuous, low density approach by Bertin 
\cite{bertin_06,bertin_09}.
Furthermore, note that $a_2$ and $b_2$, the second order moments of the distribution function $f$
appear, which, in general, could make it difficult to obtain a closed set of macroscopic equations.
Fortunately, it turns out that near the order-disorder bifurcation closure is possible; for this purpose
additional equations for $a_2$ and $b_2$ are derived by multiplying the $O(\epsilon^2)$ member
of the hierarchy Eq. (\ref{HIERARCHY_SWARM1c}) by $e_{\alpha} e_{\beta}$, and taking the angular average.
For the special case $\alpha=\beta=x$ one finds a  relation for (the rescaled) $a_2$,
\begin{equation}
\label{A2_CLOS1}
{T a_2\over 2}+{T\over 2}\circ \partial_{t_2}\rho=
\langle e_x^2 C_2\rangle-
{T\over 4}\left[ 3\partial_x w_x+\partial_y w_y\right]
-{1\over 16}\left[3\partial_x^2\rho+\partial_y^2\rho\right]
\end{equation}
A similar calculation for $\alpha=x$ and $\beta=y$ gives an expression for (the rescaled) $b_2$,
\begin{equation}
\label{B2_CLOS1}
{T\over 2} b_2=
\langle e_x e_y C_2\rangle-{T\over 4} \left[\partial_x w_y
+\partial_y w_x\right]
-{1\over 8}  \partial_x\partial_y \rho
\end{equation}
Combining both relations and working out the effect of the $T$-operators one gets
\begin{equation}
\label{A2_B2_SUM}
{T\over 2}(\partial_x a_2+\partial_y b_2)=
\partial_x\langle e_x^2 C_2\rangle+\partial_y\langle e_x e_y C_2\rangle
-{1\over 4}\nabla^2 w_x+{1\over 16}\partial_x \nabla^2 \rho
\end{equation}
which is inserted into Eq. 
(\ref{FINAL_SECOND_WX_SCAL}), yielding 
\begin{eqnarray}
\nonumber
& &R\partial_{t_2} w_x=\Gamma w_x w^2+S(w_x a_2+w_y b_2)+(\lambda-1)w_x \\
\label{A2_B2_FIN}
& &-
\partial_x\langle e_x^2 C_2\rangle-\partial_y\langle e_x e_y C_2\rangle
+{1\over 8}\nabla^2 w_x-{1 \over 48}\partial_x \nabla^2 \rho
\end{eqnarray}
Substituting the remaining moments of the collision operator $C_2$, one realizes that the higher order moments
$a_2$ and $b_2$ are still part of this evolution equation for the lower order moment $w_x$.
That means, if closure is possible at this level, there must be a way to express both $a_2$ and $b_2$ as 
functionals of $\rho$
and ${\bf w}$.
To obtain such relations, the moments of the collision operator, Eqs. (\ref{SECOND_XX_LAST}, \ref{SECOND_XY_LAST}) are inserted into 
Eqs. (\ref{A2_CLOS1}, \ref{B2_CLOS1}). This gives  
decoupled differential equations for $a_2$ and $b_2$, which are of infinite order in the time 
scale $t_0$,
\begin{eqnarray}
\label{DIFFEQ_A2}
(T-p)a_2-q(w_x^2-w_y^2)+{1\over 2}(\partial_xw_x-\partial_y w_y)-{1\over 8}
(\partial^2_x-\partial^2_y)\rho&=&0 \\
\label{DIFFEQ_B2}
(T-p)b_2-2qw_x w_y+{1\over 2}(\partial_x w_y+\partial_y w_x)-{1\over 4}
\partial_x\partial_y \rho&=&0
\end{eqnarray}
Inspection of equations similar to Eqs. (\ref{SECOND_W0},\ref{A2_CLOS1},\ref{B2_CLOS1}) but now for arbitrary $\alpha$, $\beta$,
shows that they can be expressed in a very compact and convenient way in 
terms of vectors and tensors of rank two.
This is not very surprising since the system has rotational symmetry and there should
be a rotational-invariant formulation of the macroscopic equations.
The evolution equation, Eq.~(\ref{A2_B2_FIN}), becomes
\begin{eqnarray}
\nonumber
R\,\partial_{t_2}\vec{w}&=&
(\lambda-1)\vec{w}+\Gamma w^2\vec{w}+S\siga\cdot\vec{w}
-\nabla\cdot\sig
+{1\over 8}\nabla^2\vec{w} \\
\label{FINAL_SECOND_TENSOR}
& & -{1\over 48}\nabla \nabla^2\rho\;\;\;\; \\
\label{SIGTOTDEF}
{\rm with}\;\;\;\;
\sig&=&{1\over 2}(p\siga+q\omc)
\end{eqnarray}
where $\siga$ is a traceless, symmetric tensor with $\sigma_{1,xx}=-\sigma_{1,yy}=a_2$
and $\sigma_{1,xy}=\sigma_{1,yx}=b_2$.
In this derivation, the velocity moments of $C_2$ were expressed in  terms of the flux tensor
$\sig$,
$\langle e_{\alpha} e_{\beta} C_2\rangle=\sigma_{\alpha\beta}$.
Eqs. (\ref{DIFFEQ_A2}, \ref{DIFFEQ_B2}) turn out to be the evolution equations for the components
of the tensor $\siga$,
\begin{equation}
\label{SIGMA1_DEF_TENSOR}
(T-p)\siga=q \omc-{1\over 2}\oma+{1\over 8}
\omb
\end{equation}
The $\omi$ are defined by
\begin{eqnarray}
\Omega_{1,\alpha\beta}&=&\partial_{\alpha}w_{\beta}+\partial_{\beta}w_{\alpha}
-{2\over d}\delta_{\alpha\beta}\partial_{\gamma}w_{\gamma} \\
\Omega_{2,\alpha\beta}&=&2\partial_{\alpha}\partial_{\beta}\rho
-{2\over d}\delta_{\alpha\beta}\partial^2_{\gamma}\rho \\
\Omega_{3,\alpha\beta}&=&2w_{\alpha}w_{\beta}
-{2\over d}\delta_{\alpha\beta}w^2 
\end{eqnarray}
where $d=2$ is the dimension.
$\oma$ is actually the regular stress tensor of a two-dimensional fluid.
Due to the property $\partial_{t_0}\rho=\partial^2_{t_0} w_{\alpha}=0$, and 
$\partial_{t_0}w_{\alpha}=-\partial_{\alpha}\rho/2$ (Eq. (\ref{EULER0}) in rescaled form) Eq. 
(\ref{SIGMA1_DEF_TENSOR})
can be solved by the following Ansatz,
\begin{equation}
\label{SIGMA_ANSATZ}
\siga=\sum_{i=1}^5C_i \omi
\end{equation}
which includes two new tensors 
\begin{eqnarray}
\nonumber           
\Omega_{4,\alpha\beta}&=&w_{\alpha}\partial_{\beta}\rho+w_{\beta}\partial_{\alpha}\rho
-{2\over d}\delta_{\alpha\beta}w_{\gamma}\partial_{\gamma}\rho \\
\label{OMEGA45_DEF}
\Omega_{5,\alpha\beta}&=&2(\partial_{\alpha}\rho)(\partial_{\beta}\rho)
-{2\over d}\delta_{\alpha\beta}(\partial_{\gamma}\rho)^2
\end{eqnarray}
It is easy to see that all $\omi$
are actually tensors since they can be written as direct 
products of vectors. By means of Eqs. (\ref{SIGTOTDEF},\ref{SIGMA_ANSATZ}) this also proves the previous assumption
that $\siga$ and $\sig$ are tensors.
Note, that all involved tensors are traceless and symmetric.
The Ansatz for the solution, Eq. (\ref{SIGMA_ANSATZ}), with just a finite number of terms is successful due 
to the following relations,
\begin{eqnarray}
\nonumber
\partial_{t_0} \oma &=&-{1\over 2} \omb \\
\nonumber
\partial_{t_0}\omb &=&0\\
\nonumber
\partial_{t_0}\omc&=&-\omd\\
\nonumber
\partial^2_{t_0}\omc &=&{1\over 2}\ome \\
\nonumber
\partial_{t_0}\omd&=&-{1\over 2}\ome \\
\label{OMEGA_ALG}
\partial_{t_0}\ome&=&0\,.
\end{eqnarray}
Hence, all time derivatives of order three and higher with respect to $t_0$ vanish.
The coefficients $C_i$ in Eq. (\ref{SIGMA_ANSATZ}) are found as
\begin{eqnarray}
\nonumber
C_1&=&{1\over 2(p-1)}          \\
\nonumber
C_2&=&-{p+1\over 8(1-p)^2}          \\
\nonumber
C_3&=&{q\over 1-p}          \\
\nonumber
C_4&=&{q\over (1-p)^2}          \\
\label{C5_DEF}
C_5&=&{q(1+p)\over 4(1-p)^3}
\end{eqnarray}
This concludes the solution of
the evolution equation, Eq.~(\ref{SIGMA1_DEF_TENSOR}), for the tensor $\siga$ and 
its components, Eqs.~(\ref{DIFFEQ_A2}, \ref{DIFFEQ_B2}).
This means, the higher moments $a_2$ and $b_2$ are indeed enslaved 
to the hydrodynamic fields $\rho$ and ${\bf w}$ because the solution, Eq.~(\ref{SIGMA_ANSATZ} -- \ref{C5_DEF}), 
explicitly 
shows how they are uniquely determined (at this order in $\epsilon$) by these fields and their spatial gradients.

Further analysis of Eq. (\ref{FINAL_SECOND_TENSOR}) is simplified by the observation that {\em all} terms 
can be expressed by means of the $\omi$. One finds,
\begin{eqnarray}
\label{FINAL_ALLTENSOR}
R\,\partial_{t_2}\vec{w}&=&([\lambda-1]\UN+
\Gamma \omc+S\siga)\cdot\vec{w}
-\nabla\cdot\left(\sig
-{1\over 8}\oma
+{1\over 48}\omb\right) 
\end{eqnarray}
where $\UN$ is the unity tensor.
The momentum evolution at time scale $t_2$ can then be found by the Ansatz
\begin{equation}
\label{FINAL_FORM_T2}
\partial_{t_2}\vec{w}+\nabla\cdot\HT=(u_1\UN+\QTa)\cdot \vec{w}+(u_2\UN+\QTb)\cdot\nabla\rho
\end{equation}
One finds $u_1=\lambda-1$ and $u_2=(\lambda-1)/4$.
The momentum flux tensor $\HT$ and the tensors $\QTa$, $\QTb$
are expanded as
\begin{equation}
\HT=\sum_{i=1}^5h_i\,\omi\;\;\;\;\;\;\;\
\QTa=\sum_{i=1}^5q_i\,\omi \;\;\;\;\;\;\; 
\QTb=\sum_{i=1}^5k_i\,\omi
\end{equation}
The coefficients $h_i$, $q_i$ and $k_i$ are determined by inserting Eq. (\ref{FINAL_FORM_T2})
into Eq. (\ref{FINAL_ALLTENSOR}) and using the algebra of $t_0$ derivatives, Eq. (\ref{OMEGA_ALG})
with the result,
\begin{eqnarray}
\nonumber
h_1&=&{p\over 2} C_1-{1\over 8}={1+p\over 8(p-1)} \\
\nonumber
h_2&=&{p\over 2} C_2+{p\over 8} C_1-{1\over 96}=-{p^2+10p+1\over 96(p-1)^2} \\
\nonumber
h_3&=&{p\over 2} C_3+{q\over 2}=-{q\over 2(p-1)} \\
\nonumber
h_4&=&{p\over 2} C_4+{p\over 4} C_3+{q\over 4}={q(1+p)\over 4(p-1)^2} \\
\label{H_COEFFS}
h_5&=&{p\over 2} C_5+{p\over 8} C_4+{p\over 48} C_3+{q\over 48}=-{q(p^2+10p+1)\over 48(p-1)^3}
\end{eqnarray}
and
\begin{eqnarray}
\nonumber
q_1&=& S C_1={S\over 2(p-1)} \\
\nonumber
q_2&=& S\left(C_2+{C_1\over 4}\right)=-{S\over 4(p-1)^2}\\
\nonumber
q_3&=&S C_3+\Gamma=\Gamma-{Sq\over p-1} \\
\nonumber
q_4&=&S\left(C_4+{C_3\over 2}\right)+{\Gamma\over 2}={\Gamma\over 2}-{Sq(p-3)\over 2(p-1)^2} \\
\label{Q_COEFFS}
q_5&=&S\left(C_5+{C_4\over 4}+{C_3\over 24}\right)+{\Gamma\over 24}={\Gamma\over 24}-{Sq(p^2-2p+13)\over24(p-1)^3}
\end{eqnarray}
\begin{eqnarray}
\nonumber
k_1&=&{SC_1\over 4}={S\over 8 (p-1)} \\
\nonumber
k_2&=&S\left({C_1\over 24}+{C_2\over 4} \right)=-{S (p+5)\over 96(p-1)^2}\\
\nonumber
k_3&=&{SC_3\over 4}+{\Gamma\over 4}={\Gamma\over 4}-{Sq\over 4(p-1)}\\
\nonumber
k_4&=&S\left( {C_4\over 4}+{C_3\over 12}\right)+{\Gamma\over 12}={\Gamma\over 12}-{Sq(p-4)\over 12(p-1)^2}\\
\label{K_COEFFS}
k_5&=&{q_5\over 4}+{k_4\over 4}-{q_4\over 12}-{k_3\over 12}+{q_3\over 32}=-{Sq(p+5)\over 48(p-1)^3}
\end{eqnarray}
Finally, the total momentum evolution up to cubic order in $\epsilon$ is obtained by combining
the different time derivatives and setting $\epsilon$ equal to unity,
$\partial_t\vec{w}=\partial_{t_0}\vec{w}+
\partial_{t_1}\vec{w}+
\partial_{t_2}\vec{w}+O(\epsilon^4)$,
\begin{eqnarray}
\nonumber
& &\partial_{t}\vec{w}+\nabla\cdot\HT \\
\label{FINAL_EVOLUTION_MOMENTUM}
& &=-{1\over 2}\left(1-{\lambda-1\over 2}\right)\nabla\rho+(\lambda-1)\vec{w}+
\QTa\cdot \vec{w}+\QTb\cdot\nabla\rho
\end{eqnarray}
Note, that this equation was derived at close proximity to the critical point, $\lambda\rightarrow 1$.
In the special case of constant density,$\nabla\rho=0$, 
only two of the five $\omi$ tensors, $\oma$ and $\omc$, are non-zero,
and all the density dependent coefficients are constant.
Then the Navier-Stokes-like equation for the momentum-density,
 Eq. (\ref{FINAL_EVOLUTION_MOMENTUM}), simplifies to
\begin{equation}
\label{CONSTDENSITY1}
\partial_{t}\vec{w}+h_3 \nabla\cdot\omc=(\lambda-1)\vec{w}
-h_1\nabla\cdot\oma+(q_1\oma+q_3\omc)\cdot \vec{w}
\end{equation}
This can also be written more explicitly as
\begin{eqnarray}
\nonumber
& &\partial_{t}\vec{w}+(2h_3-q_1)(\vec{w}\cdot\vec{\nabla})\vec{w}+(2h_3+q_1)(\vec{\nabla}\cdot\vec{w})\vec{w}
-\left(h_3+{q_1\over 2}\right)\vec{\nabla}\left(|\vec{w}|^2\right) \\
\label{CONSTDENSITY2}
& &=
(\lambda-1)\vec{w}-h_1 \nabla^2\vec{w}+q_3|\vec{w}|^2\vec{w}
\end{eqnarray}
The prefactor $\lambda-1$ determines the linear growth rate of the momentum, which is consistent
with the interpretation of $\lambda$ as an amplification factor in Sec.~\ref{sec:Closure}.
The growth is limited by
a cubic nonlinearity with coefficient $q_3$ which is always negative.
The coefficients $h_3$ and $q_1$ control the strength of the convective terms which are quadratic in $\vec{w}$ and contain
one gradient operator. These terms are more complicated than the one occurring in the
conventional Navier-Stokes equation because of the absence of Galilean invariance.

The coefficient $h_1$ is proportional to the shear viscosity.
More precisely, after 
returning to dimensional quantities, see 
Eqs. (\ref{NONDIM1}-\ref{NONDIM3}),
the Chapman-Enskog expansion
predicts the following kinematic shear viscosity,
\begin{equation}
\label{KIN_VISC_CHAP}
\nu^{kin}=-{v_0}^2 \tau h_1= {v_0^2 \tau\over 8}{1+p\over 1-p}\,,
\end{equation}
where $p$ is defined in Eq.~(\ref{P_DEF1}).
The viscosity is positive as expected because $|p|<1 $ for all noise values $\eta>0$.
Note, that this is only the so-called kinetic contribution to the shear viscosity because
the evaluation of the collision integrals in Sec.~{\ref{sec:Moments} was done in the limit of large mean free path
$r_0\ll v_0\tau$.
It can be shown, \cite{ihle_unpub}, that similar to dense fluids \cite{hanley_72} and 
other particle-based models \cite{gompper_08,alexander_98}, there is an additional collisional contribution, $\nu^{coll}$, 
due to collisional momentum transfer, 
\begin{equation}
\nu=\nu^{kin}+\nu^{coll}\,,
\end{equation}
This contribution is
proportional to $r_0^2/\tau$, and therefore
dominates the total viscosity $\nu$ at small mean free path.
The kinetic contribution was also derived by means of a Green-Kubo relation and
is considered in detail in 
Sec.~\ref{sec:Green_Kubo}.

\subsection{Discussion}
\label{sec:Discuss}

The original  equations for the evolution of the momentum density, 
Eqs.~(\ref{A2_B2_FIN},\ref{FINAL_SECOND_TENSOR}), still depend on higher order moments,
and require closure.
The non-trivial closure of the moment equations presented by Eqs.~(\ref{SIGMA_ANSATZ} -- \ref{C5_DEF}) 
does not simply assume
that all higher moments are zero
or constant; they still evolve in time but their dynamics is completely prescribed
by the dynamics of the lower moments -- the hydrodynamic fields $\rho$ and $\vec{w}$.
Note, that this differs from the moment closures proposed by other authors such as 
Ref. \cite{bertin_09,farrell_12,aranson_05}. These authors assume $\dot{f}_2=0$. 

The closure proposed in this paper was significantly simplified by restricting the analysis to 
the vicinity of the order-disorder threshold where $\lambda$, Eq. (\ref{LAMBDA_DEF}), is close to one.
More specifically, the scaling assumption $1-\lambda=O(\epsilon^2)$
allows one to express the time evolution of the moments of the higher order distribution functions
$f_2$ and $f_3$ in terms of gradients of the hydrodynamic fields.
This means these functions depend on time only implicitly
through their functional dependence on $f_0$ and $f_1$.
Thus, at order $O(\epsilon^3)$ and near the flocking threshold, I found that the moments $f_2$ and $f_3$ are enslaved
to $f_0$ and $f_1$, whereas even higher functions such as $f_4$ can be neglected at this order.
This results in a consistent closure of the hierarchy equations and leads to two hydrodynamic equations,
as postulated in the Toner-Tu theory \cite{toner_98}.

The transport coefficients of the final hydrodynamic equation, Eq.~(\ref{FINAL_EVOLUTION_MOMENTUM}),
have no explicit dependence on the radius of the collision circle $r_0$.
This is a consequence of the assumption of large mean free path, $v_0\tau \gg r_0$, which I used in 
the evaluation of the collisional integrals in Sec.~\ref{sec:Integral_eval}.
However, this radius will enter implicitly through $M_R$, the average number of particles per
interaction disc, $M_R=\rho\pi r_0^2$, where $\rho$ is the local particle number density.
The collisional contributions to the transport coefficients, which explicitly depend on $r_0$ and
are relevant at small mean free paths,
will be left for
future studies.

The transport coefficients were evaluated analytically in the large density limit $M\gg 1$ in Ref. \cite{ihle_15_a}
and simple expressions were given.
The time-dependent nonlinear hydrodynamic 
equations were also integrated numerically in Ref. \cite{ihle_13}, and their stability
was discussed in Ref. \cite{ihle_14_a}. 

In Sec.~\ref{sec:Green_Kubo} it is shown that the usual Green-Kubo relation for the kinetic part of the viscosity
gives a result identical to Eq.~(\ref{KIN_VISC_CHAP}), if one identifies 
the temperature $k_B T$ of the VM-``fluid'' by $v_0^2/2$
(assuming a particle mass $m=1$). 
This temperature definition is also consistent (up to terms of order $\epsilon^2$) with the pressure gradient,
derived in the Chapman-Enskog expansion because the first term on the r.h.s. of Eq. (\ref{FINAL_EVOLUTION_MOMENTUM})
can be interpreted as the gradient of an ideal gas pressure in an isothermal system. After undoing the nondimensionalization
given by Eqs. (\ref{NONDIM1}-\ref{NONDIM3}), one reads off the following pressure
\begin{equation}
\label{P_IDEAL}
p_{id}=\rho k_B T=\rho {v_0^2 \over 2} \left(1-{\lambda-1\over 2}\right)=\rho {v_0^2 \over 2}+O(\epsilon^2)
\end{equation}
This also means that the speed of sound in the VM, near the order-disorder transition where $\lambda\approx 1$,
is given by the isothermal speed of sound
\begin{equation}
\label{SOUND_SPEED}
c_T=\sqrt{k_B T}={v_0\sqrt{2}}+O(\epsilon^2)
\end{equation}
In this sense, the soliton-like density waves analyzed in Ref. \cite{ihle_13} appear to be always supersonic.

\section{Evaluation of angular integrals}
\label{sec:Integral_eval}

In order to obtain explicit expressions for the transport coefficients of the hydrodynamic equations in Sec.~\ref{sec:Moments}, 
certain moments of the collision operator, for example $\langle e_xe_yC_2\rangle$, are needed. % in Sec.~\ref{sec:Moments}.
The core part of these moments are n-dimensional integrals whose calculation for small $n$, and also asymptotically for 
very large $n$ will be presented 
in this section. The integrals are of the following type,  
\begin{equation}
I={1\over (2 \pi)^n} \int_0^{2\pi}d\alpha^{(n)}g_1(\cos{\Phi_i},\sin{\Phi_i})\,
g_2(\cos{\alpha_i},\sin{\alpha_j})
\end{equation}
where $\alpha^{(n)}=(\alpha_1,\alpha_2,...\alpha_n)$, and
$n=1,2,3,...$ is the number of actual particles in a collision cell.
$g_1$ is a product of cosine and sine of the average angle $\Phi_i$, $g_2$ is a product
of cosine and sine of the angles $\alpha_i$.
I will use the notation $K_{\alpha\beta\gamma}^{\mu\nu}$ for these integrals.
The upper index $\mu,\nu=1,2$ refers to the appearance of $\cos{\Phi_i}$ and $\sin{\Phi_i}$ 
in the function $g_1$;
the lower index
describes the product of sine and cosine of the angles $\alpha_i$.
For example, $(\mu,\nu)=(1,2)$ means $g_1=\cos{\Phi_i}\,\sin{\Phi_i}$, and $(\mu,\nu)=(1,1)$ means
$g_1=\cos{\Phi_i}\,\cos{\Phi_i}=\cos^2{\Phi_i}$.
For the lower index I will use the symbols $c$, $s$, $2c$ and $2s$ to describe the function $g_2$. 
For example, $(\alpha,\beta)=(c,2s)$ means $g_2=\cos{\alpha_1}\,\sin{2\alpha_2}$.
With this definition, one has for example:
\begin{eqnarray}
K_{c,2c}^{1,1}(n)&=&    
{1\over (2 \pi)^n} \int_0^{2\pi}d\alpha^{(n)}\cos^2{\Phi_1}\,
\cos{\alpha_1}\,\cos{2 \alpha_2}\equiv K_{c2c}^{11}\\
K_{c,c,c}^{2}(n)&=&    
{1\over (2 \pi)^n} \int_0^{2\pi}d\alpha^{(n)}\sin{\Phi_1}\,
\cos{\alpha_1}\,\cos{\alpha_2}\,\cos{\alpha_3}\equiv K_{ccc}^2\\
\end{eqnarray}
In order to determine the average angle $\Phi_1$, the local order parameter vector
${\bf L}_n=(L_{n,x},L_{n,y})$
is defined,
\begin{equation}
\label{LOCAL_ORDER1}
{\bf L}_n=\sum_{i=1}^n\,{\bf e}_i
\end{equation}
where ${\bf e}_i=v_0(\cos{\alpha_i},\sin{\alpha_i})$ is the velocity vector for particle $i$, and $v_0$
is the constant speed of a particle.
The sine and cosine of the average angle follow as usual,
$\cos{\Phi_1}=L_x/|L|$ and $\sin{\Phi_1}=L_y/|L|$.
Alternatively, the average angle can be defined directly as $\Phi_1={\rm atan}(L_y/L_x)$. 

\subsection{Calculations for finite $n$}

For $n=2$, the integrals are two-dimensional, and 
it is possible to use trigonometric sum rules to simplify the integral.
Using,
\begin{eqnarray}
{L_x(2)\over v}&=&\cos{\alpha_1}+\cos{\alpha_2}=2\cos{\alpha_1+\alpha_2\over 2}\cos{\alpha_1-\alpha_2\over 2} \\
{L_y(2)\over v}&=&\sin{\alpha_1}+\sin{\alpha_2}=2\sin{\alpha_1+\alpha_2\over 2}\cos{\alpha_1-\alpha_2\over 2} 
\end{eqnarray}
one finds that 
\begin{eqnarray}
\Phi_1&=&{\alpha_1+\alpha_2\over 2}\;\;\;{\rm for}\;\;|\alpha_1-\alpha_2|<\pi \\
\Phi_1&=&{\alpha_1+\alpha_2\over 2}+\pi\;\;\;{\rm for}\;\;|\alpha_1-\alpha_2|>\pi 
\end{eqnarray}
for $0\leq \alpha_i\leq 2 \pi$.
The integral over $\alpha_1$ and $\alpha_2$ can then be split into four parts,
\begin{eqnarray}
\nonumber
& &\int_0^{2 \pi}\,d\alpha_1\int_0^{2\pi}\,d\alpha_2...=
\int_0^{\pi}\,d\alpha_1
\left(\int_0^{\alpha_1+\pi}\,d\alpha_2...+
      \int_{\alpha_1+\pi}^{2 \pi}\,d\alpha_2\right) \\
& &+\int_{\pi}^{2\pi}\,d\alpha_1
\left(\int_{\alpha_1-\pi}^{2 \pi}\,d\alpha_2...+
      \int_{0}^{\alpha_1-\pi}\,d\alpha_2\right)
\end{eqnarray}
where in the first and third part 
$|\alpha_1-\alpha_2|<\pi$, and in the second and fourth term one has
$|\alpha_1-\alpha_2|>\pi$.
All functions under the integral are now products of sine and cosine of a linear combination $a\alpha_1+b\alpha_2$.
Therefore, all collision moments for $n=2$ can be evaluated analytically.
These exact results are given in the $(n=2)$ column of Tab.~1.
For $n\geq 3$, to my knowledge, no simple addition theorem can be used to simplify the calculations,
and the calculations become very tedious. %(beyond the abilities of Mathematica) 
I was able 
to find an analytical solution for special cases at $n=3$, see Sec.~\ref{sec:Special}, but due to the difficulty level I
 will not explore this further.
 It is faster to solve these integrals numerically, as given in Table 1. 
Furthermore, exact asymptotic solutions can be found in the limit of infinite dimensions, $n\rightarrow\infty$. By comparison to the numerical solutions one realizes 
that the asymptotic formulas are still quite accurate for $n\approx 5$ and larger.
The details of these asymptotic calculations will be presented further below.

Table 1 gives analytical and numerical results for the collision moments.
The results for $n=3,4,5$ were obtained by a straightforward numerical integration with constant intervals,
the numerical results for $n=10$ were obtained by simple Monte-Carlo integration.
It is difficult to achieve good numerical accuracy for $n>10$, since the integrand is strongly oscillating
and no regions of the angular space $(\alpha_1,...\alpha_n)$ can be neglected, hence
importance sampling methods, like the Metropolis algorithm, can not be used.
The results for $n=1$, $n=2$ and $n\rightarrow \infty$ are exact.
\vspace{0.5cm}

\begin{table}
\begin{tabular}{|r||c|c|l|l|l|l|c|}
\hline
                & $n=1$ & $n=2$    &  $n=3$    &  $n=4$     &  $n=5$     & $n=10$    & $n\rightarrow\infty$ \\ 
\hline 
\hline
$K_{c}^1$       & $1/2$ & $1/\pi$& 0.2624    & 0.2249     &  0.2008    & 0.141     &                       
 $n^{-1/2} \sqrt{\pi}/ 4  $ \\
\hline
$K_{2c}^{11}$   &  $1/4$ & 0           & 0.0544989  & 0.02700    & 0.02724    & 0.0127    &
 $ n^{-1}/8 $  \\
\hline
$K_{cs}^{12}$   &  --- & $1/8$       & 0.04888   & 0.03716    & 0.02784    & 0.0131    &
 $ n^{-1}/8 $  \\
\hline
$K_{c2c}^1$     & --- & $-1/(6\pi)$ & -0.01277  & -0.009402  & -0.005810  & -0.00143  &
 $-n^{-3/2} \sqrt{\pi}/32 $  \\
\hline
$K_{ccc}^1$     & --- &  ---    & -0.06967  & -0.03159   & -0.02074   & -0.00439  &
 $-3 n^{-3/2} \sqrt{\pi}/32 $  \\
\hline
\end{tabular}
\caption{Results for important $K$-Integrals as a function of particle number $n$. 
For details see Sec.~\ref{sec:Integral_eval}.}
\end{table}

\vspace{0.5cm}
All other similar moments are either zero or related by the ones in the table.
One finds,
\begin{eqnarray}
\nonumber
K_{css}^1&=&K_{ccc}^1/3 \\
\nonumber
K_s^2&=&K_c^1\\
\nonumber
K_{cc}^{11}&=&
-K_{ss}^{11}=
-K_{cc}^{22}=
K_{ss}^{22}=
K_{cs}^{12} \\
\nonumber
K_{2c}^{11}&=&-K_{2c}^{22}=K_{2s}^{12} \\
\nonumber
K_{c2c}^1&=&K_{s2s}^1=K_{c2s}^2=-K_{s2c}^2\\
\nonumber
K_{ccs}^2&=&K_{css}^1 \\
\label{K_RELAT2}
K_{sss}^2&=&K_{ccc}^1\,.
\end{eqnarray}
For symmetry reasons the following integrals vanish for all $n$:
\begin{eqnarray}
\nonumber
K_{ccs}^1&=&K_{sss}^1=K_{3c}^1=K_{3s}^1=K_{c2s}^1=K_{s2c}^1=0 \\
\nonumber
K_{ccc}^2&=&K_{css}^2=K_{3c}^2=K_{3s}^2=K_{c2c}^2=K_{s2s}^2=0 \\
\nonumber
K_{2c}^1&=&K_{cc}^1=K_{ss}^1=K_{cs}^1=K_{c2c}^2=K_{s2s}^2=0\\
\nonumber
K_{2c}^{12}&=&K_{cs}^{11}=K_{2s}^{11}=K_{2s}^{22}=K_{cc}^{12}=K_{ss}^{12}=0 \\
\nonumber
K_{3c}^{11}&=&K_{3s}^{12}=K_{3c}^{12}=K_{3c}^{22}=K_{3s}^{22}=0 \\
\nonumber
K_{ccc}^{11}&=&K_{c2c}^{11}=K_{css}^{11}=K_{sss}^{11}=K_{ccs}^{11}=0 \\
K_{ccc}^{12}&=&K_{scc}^{12}=K_{ssc}^{12}=K_{sss}^{12}=0
\end{eqnarray}
Higher moments, such as $K_{c3c}^{11}$ can be non-zero (-0.002811 for $n=3$) but occur at higher order 
than $\epsilon^3$ in the Chapman-Enskog expansion.
Note, that $K_{cs}^{11}$ has a non-monotonic behavior, and more interestingly, exactly vanishes for binary collisions, 
$n=2$. 

\subsection{Calculations for $n\rightarrow \infty$}

The main idea of how to obtain the asymptotic behavior is based on an analogy to a random walk or a Gaussian chain:
The individual velocity vectors ${\bf e}_i$ can be seen as segments of fixed length $v_0$ forming a random chain.
The local order parameter vector ${\bf L}_n$, see Eq. (\ref{LOCAL_ORDER1}), 
translates then into the end-to-end vector for this chain of $n$ segments.
It is well known that in the limit of infinite chain length, the probability density to realize 
a certain end-to-end vector 
is Gaussian,
\begin{equation}
\label{GAUSS}
p({\bf L})={1\over \pi v_0^2 n}{\rm e}^{-L^2/(v_0^2 n)}\,,
\end{equation}
due to the central limit theorem.
In particular, the average chain length $\langle |{\bf L}|\rangle$ is of order $v_0 n^{1/2}$, i.e.
is large for large $n$. This means $1/L$ is small for most chain realizations and will be used
as an expansion parameter in the following.
Eq. (\ref{GAUSS}) contains the obvious fact that all directions of the vector ${\bf L}$ are equally probable, i.e. 
the probability density for the angle $\beta$ of the order parameter vector is the same for all angles
between zero and $2\pi$. 

\vspace{0.5cm}

\subsubsection{One-angle calculations}

I will start by considering integrals where the integrand contains only {\em one} angular argument, $\alpha_1$,
in addition to functions of the average angle $\Phi_1$.
The order parameter vector can be split up in the following way:
\begin{equation}
\label{SPLIT1}
{\bf L}_n={\bf L}_{n-1}+{\bf e}_1
\end{equation}
with the vector ${\bf e}_1=v_0(\cos{\alpha_1}, \sin{\alpha_1})$.
The components of ${\bf L}_{n-1}$ can be expressed in terms of the angle $\beta$ and length $L_{n-1}$ of this vector,
$L_{n-1,x}=c\,L_{n-1}$, $L_{n-1,y}=s\,L_{n-1}$, where $c=\cos{\beta}$ and $s=\sin{\beta}$.
As a result, the cosine of the average angle can be written as,
\begin{eqnarray}
\label{SPLIT2}
\cos{\Phi_1}&=&{L_{n-1} c+v_0 c_1\over \sqrt{
(L_{n-1} c+v_0 c_1)^2
+(L_{n-1} s+v_0 s_1)^2}}
\end{eqnarray}
with $c_1=\cos{\alpha_1}$, $s_1=\sin{\alpha_1}$.
A similar expression follows for $\sin{\Phi_1}$.
Now the $n-1$ integrations over the angles $\alpha_2,\alpha_3...\alpha_n$ can be replaced 
by an integration over all possible lengths $L_{n-1}$, which are between zero and $(n-1)v_0$ and over the orientation angle $\beta$ of the end-to-end vector.
In the limit of infinite dimensions, $n\rightarrow \infty$, the upper limit of the integral over all possible 
lengths goes to infinity and we can use the simple Gaussian expressions, Eq. (\ref{GAUSS}) for the probability density.
Finally, for example $K_{2c}^{11}$ becomes,
\begin{eqnarray}
\nonumber    
& &K_{2c}^{11}={1\over 2\pi}\int_0^{2\pi}\,d\beta \int_0^{\infty}p(L_{n-1})L_{n-1}\,dL_{n-1} \\
\label{K2C_1}
& &\times\int_0^{2 \pi}\,d\alpha_1 
\cos{2\alpha_1}{ (L_{n-1}c+v_0 c_1)^2\over (L_{n-1}c+v_0 c_1)^2+(L_{n-1}s+v_0s_1)^2}
\qquad\;n\rightarrow \infty
\end{eqnarray}
where $p(L_{n-1})$ is given in Eq. (\ref{GAUSS}).
After introducing the dimensionless variable $L=L_{n-1}/v_0$
this can be written by means of the function $h(L)$ as
\begin{eqnarray}
\label{K211_2}
K_{2c}^{11}&=&{1\over \pi(n-1)} \int_0^{\infty}h(L)L{\rm e}^{-L^2/(n-1)}\,dL\,,\;\;\;{\rm with} \\
h(L)&=&{1\over 2\pi}\int_0^{2\pi}\,d\beta \int_0^{2 \pi}\,d\alpha_1 
\cos{2\alpha_1}{ (Lc+c_1)^2\over (Lc+c_1)^2+(Ls+s_1)^2}
\end{eqnarray}
I found that $h(L)$ has a very simple structure with a  cusp at $L=1$:
$h(L)={\pi\over 2}(1-L^2)$ for $0\leq L\leq 1$ and $h(L)=0$ for $L\geq 1$,
which can be easily checked for $L=0$ and $L=1$ but is nontrivial to prove for arbitrary $L$.
This simple quadratic expression agreed perfectly with a numerical evaluation of $h(L)$.
Since $h=0$ for $L\geq 1$, the upper limit in Eq. (\ref{K211_2}) can 
be reduced to one, and in the limit of infinite $n$, the exponential factor ${\rm exp(-L^2/n)}$ goes to one
for all $L$ in the integration range, and $n-1$ is replaced by $n$.
The asymptotic result is therefore:
\begin{equation}
\label{K2C11_AS}
K_{2c}^{11}={1\over \pi n} \int_0^1 {\pi\over 2} L(1-L^2)\,dL={1\over 8n} 
\end{equation}
Similar calculations were performed for other integrals, such as $K_{2s}^{12}$, leading to the same
asymptotic behavior $\sim n^{-1}$.
Simulations showed that the integral $K_{c}^1$ however has a weaker decay than $\sim n^{-1}$ 
for $n\rightarrow \infty$ and will therefore be analyzed in the following.
Applying the same transformation from individual angular coordinates, $(\alpha_2, \alpha_3)$
to $(\beta, L)$ with $L=L_{n-1}/v_0$, one finds,
\begin{eqnarray}
\label{KC1}
K_{c}^{1}&=&{1\over 2\pi}\int_0^{2\pi}\,d\beta \int_0^{\infty}p(L)L\,dL\int_0^{2 \pi}\,d\alpha_1 
\cos{\alpha_1} \cos{\Phi_1}\,\;
\qquad\;n\rightarrow \infty
\end{eqnarray}
where $\cos{\Phi_1}$ is given in Eq.~(\ref{SPLIT2}).
For $L>1$ the square root is expanded in terms of order $(1/L)^m$. 
This gives
\begin{eqnarray}
\nonumber    
& &{1\over \sqrt{(L_{n-1}c+v_0 c_1)^2+(L_{n-1}s+v_0s_1)^2}} \\
\label{KC1A}
& &={1\over \sqrt{L^2+1}}
\left(1-{L\over L^2+1} (c c_1+s s_1)+O(L^{-2}) \right)
\end{eqnarray}
The integral over $L$ is split into two parts with $L_c$ fixed but much larger than one,
\begin{equation}
\label{KC1B}
K_{c}^{1}=\int_0^{L_c}\,dL\,...
+\int_{L_c}^{\infty}\,dL\,...
\end{equation}
because in the second term it is justified to use the expansion of the square root.
The integration over the angular coordinates, $\beta$ and $\alpha_1$ can now be performed easily for this term.
One obtains,
\begin{eqnarray}
\nonumber
& &K_{c}^{1}={1\over 2\pi}\int_0^{2\pi}\,d\beta \int_0^{L_c}p(L)L\,dL\int_0^{2 \pi}\,d\alpha_1 
\cos{\alpha_1}
\cos{\Phi_1} \\
\label{KC1B}
& &
+{\pi\over 2} \int_{L_c}^{\infty}{L\,dL\,p(L)\over\sqrt{L^2+1}}\left[2-{L^2\over L^2+1}\right]
\end{eqnarray}
In the limit $n\rightarrow \infty$, $p(L)$ is proportional $n^{-1}$;
$\cos{\Phi_1}$ is always of order one, hence one sees that the first integral decays at least as quickly as $n^{-1}$. On the other hand, as it turns out, the second integral decays slower than this, thus the leading 
asymptotic behavior of $K_c^1$ is given by the second integral. % from $L_c$ to $\infty$. 
In particular, after inserting $p(L)$, replacing $n-1$ by $n$ and the substitution $L=xn^{1/2}$, one finds
\begin{equation}
\label{KC1C}
K_{c}^{1}\sim
{1\over 2 n^{1/2}}\,\int_{L_cn^{-1/2}}^{\infty}\,{x\,dx\over \sqrt{x^2+n^{-1}}}\left[2-{x^2\over x^2+n^{-1}}\right]
{\rm e}^{-x^2}
\end{equation}
In the limit $n\rightarrow \infty$, this turns into a simple Gaussian integral,
\begin{equation}
\label{KC1D}
K_{c}^{1}\sim
{1\over 2 n^{1/2}}\,\int_0^{\infty}\,{\rm e}^{-x^2}\,dx={1\over 4}\sqrt{\pi\over n}
\end{equation}

\subsubsection{Two-angle calculations}

Here I will consider integrals where the integrand contains {\em two} angular arguments, $\alpha_1$ and $\alpha_2$.
Examples are, $K_{cs}^{12}$, $K_{cc}^{11}$ and $K_{ss}^{11}$.
Similar to Eq. (\ref{SPLIT1}), the end-to-end vector ${\bf L}$ is split up into three contributions
\begin{equation}
\label{SPLIT3}
{\bf L}_n={\bf L}_{n-2}+{\bf e}_1+{\bf e}_2\,,
\end{equation}
and the integration over the $n-2$ angles $\alpha_3,\alpha_4,...\alpha_n$ is replaced by an integration
over the length and the angle $\beta$ of the vector ${\bf L}_{n-2}$.
Analogous to the previous section, in the limit of $n\rightarrow \infty$ one finds  for example
\begin{eqnarray}
\nonumber      
K_{cs}^{12}&=&{1\over \pi(n-2)} \int_0^{\infty}g(L)L{\rm e}^{-L^2/(n-2)}\,dL\,,\;\;\;{\rm with} \\
\nonumber
g(L)&=&{1\over (2\pi)^2}\int_0^{2\pi}\,d\beta \int_0^{2 \pi}\,d\alpha_1  \\
\label{KCS12_1}
& &
\times\int_0^{2 \pi}\,d\alpha_2
\sin{\alpha_1} \cos{\alpha_2}{(Ls+s_1+s_2) (Lc+c_1+c_2)\over (Lc+c_1+c_2)^2+(Ls+s_1+s_2)^2}
\end{eqnarray}
It turns out that $g(L)=0$ for $L\geq 2$ and that 
\begin{equation}
{1\over \pi}\int_0^2L h(L)\,dL={1\over 8}
\end{equation}
Because of that, the asymptotic behavior is identical to the one of $K_{2c}^{11}$, and one has
\begin{equation}
\label{KCS12_AS}
K_{cs}^{12}={1\over 8n}\,,\qquad\;{\rm for}\;n\rightarrow \infty\,. 
\end{equation}
A similar analysis leads to
\begin{eqnarray}
\nonumber          
K_{cc}^{11}&=&{1\over 8n} \\ 
\label{KC_REST_AS}
K_{ss}^{11}&=&-{1\over 8n}\,,\qquad\;{\rm for}\;n\rightarrow \infty\,. 
\end{eqnarray}

\subsubsection{Three-angle calculations}

The end-to-end vector ${\bf L}$ is split up into four contributions
\begin{equation}
\label{SPLIT3}
{\bf L}_n={\bf L}_{n-3}+{\bf e}_1+{\bf e}_2+{\bf e}_3\,,
\end{equation}
and the integration over the $n-3$ angles $\alpha_4,\alpha_5,...\alpha_n$ is replaced by an integration
over the length and direction of ${\bf L}_{n-3}$.
Asymptotically one finds
\begin{eqnarray}
\nonumber      
K_{ccc}^{1}&=&{1\over \pi(n-3)} \int_0^{\infty}q(L)L{\rm e}^{-L^2/(n-3)}\,dL\,,\;\;\;{\rm with} \\
\nonumber
q(L)&=&{1\over (2\pi)^3}\int_0^{2\pi}\,d\beta \int_0^{2 \pi}\,d\alpha_1 
\int_0^{2 \pi}\,d\alpha_2
\int_0^{2 \pi}\,d\alpha_3
\\
\label{KCCC1_1}
& & \times
c_1 c_2 c_3 {(Lc+c_1+c_2+c_3\over \sqrt{(Lc+c_1+c_2+c_3)^2+(Ls+s_1+s_2+s_3)^2}}
\end{eqnarray}
This time $q(L)$ does not vanish exactly above a certain finite value of $L$.
In order to proceed, for $L>L_c\approx 3$, the square root can be expanded in a convergent series 
\begin{eqnarray}
\nonumber        
& &{1\over \sqrt{(Lc+c_1+c_2+c_3)^2+(Ls+s_1+s_2+s_3)^2}}
={1\over \sqrt{L^2+3}} {1\over \sqrt{1+g}} \\
\nonumber        
& &\approx
{1\over \sqrt{L^2+3}} (1-{1\over 2}g+{3\over 8}g^2-{15\over 48}g^3+... \\
\nonumber
& & g={2L\over L^2+3}\left[ c c_1+c c_2+c c_3+s s_1+s s_2+s s_3 \right] \\
\label{KCCC1_2}
& &
+{2\over L^2+3}\left[ s_1 s_2+s_1 s_3+s_2 s_3+c_1 c_2 +c_1 c_3+c_2 c_3\right]
\end{eqnarray}
This expansion must be done to third order in $g$ to capture the leading asymptotic behavior of the integral,
\begin{eqnarray}
\nonumber
q(L)&=&{\phi_0\over L^3}+O(L^{-4})\,\qquad\;{\rm with} \\ 
\label{LEAD_Q}
\phi_0&=&{3\pi\over 32 }\,,\qquad{\rm for}\;L>L_c
\end{eqnarray}
For $0\leq L\leq L_c$ the expansion breaks down, the function $q(L)$ varies strongly, shows one minimum and one maximum, and the contribution
to the final integral over $L$ cannot be neglected.
However, it turns out that the total integral vanishes, which was confirmed by numerical integration:
\begin{equation}
\label{Q_ZERO}
\int_0^{\infty}L q(L)\,dL=0
\end{equation}
This condition is sufficient to obtain an exact formula for the asymptotic behavior of $K_{ccc}^1$ and 
similar integrals:
Let us split Eq. (\ref{KCCC1_1}) into two parts:
\begin{equation}
K_{ccc}^{1}={1\over \pi(n-3)} \int_0^{L_s} q(L)L{\rm e}^{-L^2/(n-3)}\,dL
+\int_{L_s}^{\infty} q(L)L{\rm e}^{-L^2/(n-3)}\,dL
\end{equation}
where $L_s$ is a fixed number, $L_s\gg L_c$. 
In the first term, the exponential can be replaced by one for large $n\gg L_s^2$.
In the second term, we can safely substitute the asymptotic expression for $q(L)$ from Eq. (\ref{LEAD_Q}),
and
$n-3$ can be replaced by $n$.
\begin{equation}
\label{NOCHMAL1}
K_{ccc}^{1}={1\over \pi n} \int_0^{L_s} q(L)L\,dL
+\int_{L_s}^{\infty} {\phi_0\over L^2} {\rm e}^{-L^2/(n-3)}\,dL
\end{equation}
with $\phi_0=3\pi/32$.
Now, the first integral can be written as
\begin{eqnarray}
\nonumber
& &\int_0^{L_s} q(L)L\,dL=\int_0^{\infty} q(L) L-\int_{L_s}^{\infty} q(L) L\,dL \\
\label{NOCHMAL2}
& &
=-\int_{L_s}^{\infty} q(L) L\,dL=-\phi_0\int_{L_s}^{\infty}\,{dL\over L^2}
\end{eqnarray}
where I made use of property (\ref{Q_ZERO}) and inserted the asymptotic expression for $q(L)$, which is justified
for the range of large L in this integral.
Combining Eqs.~(\ref{NOCHMAL1},\ref{NOCHMAL2}), one obtains,
\begin{equation}
K_{ccc}^{1}={\phi_0\over \pi n} \int_{L_s}^{\infty} {{\rm e}^{-L^2/n}-1\over L^2}\,dL
\end{equation}
At this point, the substitution $x=L/\sqrt{n}$ is made,
\begin{equation}
K_{ccc}^{1}={\phi_0\over \pi n^{3/2} } \int_{L_s/\sqrt{n}}^{\infty} {{\rm e}^{-x^2}-1\over x^2}\,dx
\end{equation}
Since $L_s$ is a fixed number, the lower limit of this integral goes to zero for $n\rightarrow \infty$
and one has
\begin{equation}
K_{ccc}^{1}=-{\phi_0\over \pi n^{3/2} } \int_0^{\infty} {1-{\rm e}^{-x^2}\over x^2}\,dx
\end{equation}
The remaining integral can be solved exactly.
The final asymptotic result is
\begin{equation}
K_{ccc}^{1}=-{3\sqrt{\pi}\over 32}{1\over n^{3/2}}
\end{equation}

\subsection{A special case at $n=3$}
\label{sec:Special}

It is possible to obtain 
an analytical solution for certain $K$-integrals at $n=3$, for example for $K_{2c}^{11}$.
First, I use the decomposition from Eq. (\ref{SPLIT1}), ${\bf L}_3={\bf L}_2+{\bf e}_1$.
The probability density $p(\beta,L_2)$ to find the vector ${\bf L}_2$ at certain angle $\beta$ and certain length $L$ is now far from
being exponential and actually is zero for all $L>2v_0$.
However, ${\bf L}_2$ is only composed of two segments, each of length $v_0$; and for given angle $\beta$
and length $L_2$, there is only two ways to add up two segments to create the vector. 
If the angles of the corresponding segments are given by $\alpha_2$ and $\alpha_3$, respectively,
these two solutions are given by:
$(\alpha_2,\alpha_3)=(\beta+\gamma,\beta-\gamma)$ or $(\beta-\gamma,\beta+\gamma)$ where $\gamma$
involves the inverse cosine taken in the first quadrant, $\gamma=\cos^{-1}{(L/2v_0)}$.
Calculating the Jacobian for the transformation from the variables $(\alpha_2,\alpha_3)$ to the new variables,
$(\beta,L)$ and taking into account that there is two possible ways to construct the same vector, the 
integral over the two segment orientations can be written as:
\begin{equation}
\int_0^{2\pi}\,d\alpha_2\int_0^{2\pi}\,d\alpha_3\,...=
\int_0^{2\pi}\,d\beta\int_0^{2v_0}{2\over v_0\sqrt{1-L^2/(4 v_0^2)}}\,dL_2\,...
\end{equation}
That means the probability density to find vector ${\bf L}_2$ at lengths between $L$ and $L+dL$ and angles
between $\beta$ and $\beta+d\beta$ is given by
\begin{equation}
p(\beta,L_2)={1\over 2\pi^2 v_0\sqrt{1-L^2/(4 v_0^2)}}
\end{equation}
for $0\leq L_2<2v_0$ and is zero otherwise.
Now, I can rewrite expression (\ref{K2C_1}) in terms of the new probability density: 
Using the dimensionless variable $L=L_2/v_0$, I find,
\begin{eqnarray}
\nonumber
& &K_{2c}^{11}={1\over (2\pi)^3}\int_0^{2\pi}\,d\alpha_1
\int_0^{2\pi}\,d\alpha_2
\int_0^{2\pi}\,d\alpha_3\,\cos^2{\Phi_1}\,\cos{2\alpha_1} \\
\label{SPECIAL_N3}
& &={1\over 2\pi^2}\int_0^2{h(L)\,dL\over \sqrt{1-L^2/4}}
\,.
\end{eqnarray}
Note that $h(L)=\pi(1-L^2)/2$, thus this integral can be solved analytically,
\begin{equation}
K_{2c}^{11}(n=3)={1\over 4\pi}\int_0^{1}{1-L^2\over \sqrt{1-{L^2\over 4}}}\,dL=
{\sqrt{3}\over 4\pi}-{1\over 12}=0.05449889...
\end{equation}
This result agrees perfectly with the numerical result from Table 1.

\section{Consistency checks via Green-Kubo relations}
\label{sec:Green_Kubo}

In this section, velocity and stress auto-correlation functions will be evaluated analytically
in the Molecular Chaos approximation. These expressions can be useful for consistency checks in agent-based
simulations of the VM. 
In addition, these auto-correlation functions can be  used in Green-Kubo relations.
These relations make neither use of the main kinetic equation (\ref{ENSKOG_MAIN}) nor of the Chapman-Enskog expansion. 
Therefore, they provide a fast,
alternative route to important transport coefficients such as the shear viscosity of a VM-fluid. 
Green-Kubo relations are usually used for fluids in thermal equilibrium.
It will be shown below that they also give accurate and consistent results for a non-equilibrium
``fluid'' such as the Vicsek-model.

Consider the velocity autocorrelation of the focal particle $i=1$,
\begin{equation}
G(n\tau)=\langle v_{1x}(0)v_{1x}(n\tau)\rangle
\end{equation}
Hence, 
$G(0)=\langle v_{1x}^2\rangle=v_0^2/2$. 
For a time lag of one time step, $n=1$, we have
\begin{equation}
G(\tau)
=v_0^2 \langle \cos\alpha_1 \cos{(\Phi_1+\xi_1)}\rangle 
\end{equation}
where $\xi_1$ is a random angle which is equally distributed in the interval $[-\eta/2,+\eta/2]$, and
$\Phi_1(\{\alpha_j\})$ is the average angle. The precollisional angles are denoted by $\alpha_j$.
Applying trigonometric identities and averaging over the noise term $\xi_1$ gives
\begin{equation}
\label{HAT_G1}
\hat{G}(\tau)\equiv{G(\tau)\over k_B T}=4 {\sin{(\eta/2)} \over \eta} \langle \cos\alpha_1 \cos{\Phi_1}\rangle_c 
\end{equation}
Here, I divided by $v_0^2/2=k_B T$ because the kinetic temperature in a two-dimensional system is defined by
$\langle \vec{v}^2 \rangle=2\,k_B T$ for particles with mass $m=1$. 
This temperature definition is also consistent (up to terms of order $\epsilon^2$) with the pressure
derived in the Chapman-Enskog expansion,
see discussion in Sec.~\ref{sec:Discuss}.
The brackets with the subscript ``c'' in Eq.~(\ref{HAT_G1}) denote averages not only over the angles of all $n$ particles which are contained
in the collision circle of particle $1$ but also over the number $n$ of these particles itself (including particle $1$),
\begin{equation}
\langle\ldots\rangle_c\equiv \sum_{n=1}^N {p_n\over (2\pi)^n} \prod_{i=1}^n \int_0^{2\pi} \,d\alpha_i \ldots  
\end{equation}
Assuming Molecular Chaos, the number of particles in a circle around a focal particle is Poisson-distributed
with $p_n=M_R^{n-1}/(n-1)!$.
Thus, one obtains 
\begin{equation}
\hat{G}(\tau)={4\over \eta} \sin{\eta\over 2} \langle \cos\alpha_i \cos{\Phi_i}\rangle_c=
{4\over \eta} \sin{\eta\over 2}\sum_{n=1}^{N}{\rm e}^{-M_R}
{M_R^{n-1}\over (n-1)!} K_C^1(n)
\end{equation}
where the integrals $K_C^1(n)$ are given in Table 1.

The asymptotic expression for large $M_R$ can be obtained by the procedure outlined in 
Eqs.~({\ref{POIS_SIMP}--\ref{EXPAND_H})
with the result
\begin{equation}
\hat{G}(\tau)\approx {4\over \eta} \sin{\eta\over 2} 
K_C^1(M+1)\approx {1\over \eta} \sin{\eta\over 2} \sqrt{\pi\over M+1}
\end{equation}
For $\eta=2\pi$ the velocity correlation is exactly zero. This is expected because the new velocity directions are chosen 
completely random in this limit.
The correlation is also predicted to vanish in the limit $M_R\rightarrow \infty$.
This makes sense because in this limit the focal particle has a vanishing impact on the average angle $\Phi_1$:
the focal particle is 
``overpowered'' by the infinitely many other particles in its collision circle. 

The
Green-Kubo relation
for the self-diffusion coefficient in a system with discrete time step is given by
\begin{equation}
D=\tau {\sum_{n=0}^{\infty}}' G(n \tau)
\end{equation}
where the prime on the sum indicates that the $n=0$ term is weighted by the factor $1/2$,
see Ref. \cite{tuzel_03}.
Assuming molecular chaos leads to a geometric series for the higher order time correlations function, 
$\hat{G}(n\tau)= (\hat{G}(\tau))^n$, and the self-diffusion coefficient follows as 
\begin{equation}
\label{DIFF_COEFF_FIN}
D={\tau v_0^2\over 4}\; {1+\hat{G}(\tau)\over 1-\hat{G}(\tau)}
\end{equation}
This formula is consistent with the expectation for a completely uncorrelated random walk in two dimensions with
fixed step size
$l=v\,\tau$ which is realized for $\eta=2\pi$,
\begin{equation}
D_{random}={\tau v^2\over 4}
\end{equation}

To derive the shear viscosity, the autocorrelation function of the $xy$-component of the kinetic stress tensor
$\sigma^{kin}_{xy}=\sum_{i=1}^N v_{ix} v_{iy}$,
\begin{equation}
\label{GSHEAR1}
G^{kin}(n\tau)=\langle \sigma^{kin}_{xy}(0) \sigma^{kin}_{xy}(n\tau)\rangle
\end{equation}
is evaluated in a similar way. 
Already at the equal time level, $n=0$, there is a difference to a regular fluid.
This is because in the VM the x- and y-component of a given particle's velocity are strongly correlated,
$v_y=\sqrt{v_0^2-v_x^2}$.
As a result, $G^{kin}(0)$ is only half as big as in a regular fluid,
\begin{eqnarray}
\nonumber
G^{kin}(0)&=&
\sum_{i=1}^N\sum_{j=1}^N
\langle v_{ix}(0) v_{iy}(0) v_{jx}(0) v_{jy}(0) \rangle \\
\nonumber
&=& N\langle v_{1x}^2(0) v_{1y}^2(0) \rangle=N \left( \langle v_{1x}^2 \rangle- \langle v_{1x}^4 \rangle\right) \\
\label{GSHEAR2}
&=&{Nv_0^4\over 8}={N\, (k_B T)^2\over 2}
\end{eqnarray}
For $n=1$ one has,
\begin{eqnarray}
G^{kin}(\tau)&=&\sum_{i=1}^N\sum_{j=1}^N
\langle v_{ix}(0) v_{iy}(0) v_{jx}(\tau) v_{jy}(\tau) \rangle \\
\label{GSHEAR3}
&=&N \sum_{j=1}^N
v_0^4\langle \cos\alpha_1 \sin\alpha_1 \cos{(\Phi_j+\xi_j)} \sin{(\Phi_j+\xi_j)} \rangle
\end{eqnarray}
If molecular chaos is assumed,
only those $n$ particles that are in the collision circle around the focal particle $i=1$, contribute to the sum
over $j$ on the r.h.s. of Eq. (\ref{GSHEAR3}). 
All those particles contribute to the average angle $\Phi_1$, i.e. $\Phi_j=\Phi_1$, and one finds
\begin{eqnarray}
G^{kin}(\tau)&=&
N v_0^4\langle n \cos\alpha_1 \sin\alpha_1 \cos{\Phi_1} \sin{\Phi_1} \rangle_c
\langle \cos{2\xi_1}\rangle
\end{eqnarray}
The noise average gives
$\langle \cos{2\xi_1}\rangle=\sin{(\eta)}/\eta$. Averaging over the Poisson-distributed 
particle number fluctuations
and the pre-collisional angles $\alpha_j$ yields,
\begin{eqnarray}
& &\langle n \cos\alpha_1 \sin\alpha_1 \cos{\Phi_1} \sin{\Phi_1} \rangle_c 
=
{1\over 2}
\langle n \sin{(2\alpha_1)} \cos{\Phi_1} \sin{\Phi_1} \rangle_c \\
& &=
\sum_{n=1}^{N}{\rm e}^{-M_R}
{n\,M_R^{n-1}\over 2 (n-1)!} K_{2s}^{12}(n)
\end{eqnarray}
According to Eq.~(\ref{K_RELAT2}), $K_{2s}^{12}=K_{2c}^{11}$,
and it follows that
\begin{eqnarray}
G^{kin}(\tau)&=&
{N v_0^4\over 2} 
\sum_{n=1}^{N}{\rm e}^{-M_R}
{n^2\,M_R^{n-1}\over n!} K_{2c}^{11}(n)
\end{eqnarray}
Replacing $v_0^2/2$ by $k_B T$ and using the auxiliary variable $p$ from Eq.~({\ref{P_DEF1})
gives the final result for the kinetic stress correlations,
\begin{equation}
\label{FINAL_GKIN}
G^{kin}(\tau)= {1\over 2} (k_B T)^2 N\, p
\end{equation}
Due to Molecular Chaos, the temporal correlations decay as a geometric series,
\begin{equation}
\label{P_INTERPRET}
{G^{kin}(\tau)\over G^{kin}(0)}=
{G^{kin}([n+1]\tau)\over G^{kin}(n\tau)}=const=p
\end{equation}
where the constant ratio was found by using Eqs.~(\ref{GSHEAR2}) and (\ref{FINAL_GKIN}).
The kinetic part of the shear viscosity is then obtained by means of the usual Green-Kubo relation
for a system with discrete time dynamics, see for example Ref.\cite{tuzel_03,ihle_05},
\begin{eqnarray}
\nonumber
\nu^{kin}&=& {\tau\over N k_B T}{\sum_{n=0}^{\infty}}' G^{kin}(n \tau)=
{\tau\over N k_B T} G^{kin}(0) \left( {1\over 2}+\sum_{k=1}^\infty p^k\right)\\
\label{FINAL_GK_VIS}
&=&
{\tau k_B T\over 4} {1+p\over 1-p}={v_0^2 \tau\over 8} {1+p\over 1-p}
\end{eqnarray}
The integral $K^{11}_{2c}$ is given in Table 1, and the asymptotic expressions for $p$ are provided in Eqs.~(\ref{SECOND_AP_P})
 and (\ref{SECOND_FIN_P}). 
The result for $\nu^{kin}$, Eq.~(\ref{FINAL_GK_VIS}), agrees perfectly with the shear viscosity $-v_0^2 \tau h_1$, 
Eqs.~(\ref{H_COEFFS}) and (\ref{KIN_VISC_CHAP}), 
which was derived by a completely different approach -- a
kinetic theory based on an Enskog-like equation and a subsequent Chapman-Enskog expansion.
This agreement provides a non-trivial consistency test for the kinetic theory presented in this paper.

\section{Conclusion}
Macroscopic evolution equations for interacting many-body systems do not just ``emerge''; they follow from microscopic laws.
However, it is often difficult
to quantitatively establish this link. This is especially the case for open systems which cannot be described by
a Hamiltonian and which might have genuine multi-particle
interactions that are not pairwise additive. 
Therefore, the general form of the macroscopic
equations is often obtained by symmetry arguments. This 
can lead to hydrodynamic equations
with many unknown parameters.
The Vicsek-model (VM) \cite{vicsek_95} is a well-known example for this kind of open systems.
It is one of the simplest models to study collective motion of self-driven particles. 
Here, 
I show how
the macroscopic transport equations can be systematically derived from the microscopic interaction rules of the 
standard VM.
Whereas most of the results of this derivation have already been published and briefly discussed \cite{ihle_11,ihle_14_a,ihle_15_a}, 
in this paper, the details of the
used analytical techniques together with additional insights are presented.
I set up the exact evolution equation for the N-particle probability distribution and show how it can be reduced to
an Enskog-like kinetic equation for the one-particle density by means of the molecular chaos approximation.
No linearization or single-relaxation time approximation of the collision operator are needed, and the particle density does not have to be small.
A non-standard Chapman-Enskog expansion of the kinetic equation in the formal ordering parameter $\epsilon$, and a self-consistent closure 
of the infinite hierarchy of moment equations is proposed.
This procedure involves an expansion in spatial and temporal gradients, and contains
a fast time scale. By means of the Chapman-Enskog expansion, hydrodynamic equations for the density and momentum density
are derived and all transport coefficients that are relevant up to third order in $\epsilon$ are given.
The transport coefficients depend on special $n$-dimensional integrals which I call $K$-integrals.
It is shown, how these integrals can be analytically evaluated for $n=1,2,3$ and for $n\rightarrow \infty$.

Apart from explaining the elaborate analytical techniques, the main results of this paper are the following:
 (i) the hydrodynamic equation for the momentum density, Eq. (\ref{FINAL_EVOLUTION_MOMENTUM}) and the corresponding
transport coefficients in Eqs.~(\ref{H_COEFFS}--\ref{K_COEFFS}), (ii) the insight that even in a nonequilibrium model with correlated components of the particle velocity such as the VM, the expression for the shear viscosity, 
Eqs.~(\ref{KIN_VISC_CHAP},\ref{FINAL_GK_VIS}),
can be obtained by either a Green-Kubo relation or the Chapman-Enskog expansion, (iii) the analytical and numerical results for
the angular integrals in Table 1, and (iv) a formula for the self-diffusion coefficient of the VM, Eq. (\ref{DIFF_COEFF_FIN}).

\section{Acknowledgments}
Support
from the National Science Foundation under grant No.
DMR-0706017
is gratefully acknowledged.
I would also like to thank A. Nikoubashman 
for discussions that initiated the evaluation of the Green-Kubo relations.
\appendix

\section{Linear stability analysis}
\label{app:Stability}

It is interesting to investigate the stability of a homogeneous, globally ordered state 
against small perturbations of density and order.
The results of such an analysis for the standard Vicsek-model have been briefly presented in Refs. \cite{ihle_11,ihle_14_a},
and in Fig.~10 of Ref. \cite{chou_12} were compared to another version of the VM.
In Ref. \cite{bertin_09} a similar analysis has been performed for a Vicsek-like model with binary interactions and continuous time dynamics.
Furthermore, in Ref. \cite{ihle_15_a} a detailed calculation was given for the special case of pure longitudinal perturbations
in the large density limit, $M\gg 1$. The main insight was, that 
a long wave length instability occurs right at the onset of collective order. The growth of the perturbations finally leads 
to the formation of large density waves that show hysteresis, and provide a mean-field mechanism to modify the character of
the order-disorder transition from continuous to discontinuous, see Ref. \cite{ihle_13}. 
This seems to be a generic result which occurs in
VM-like models with a coupling between local density 
and order, and is consistent with earlier results on a Vicsek-like model with binary interactions, Refs. \cite{bertin_06,bertin_09}.
VM-like models without a coupling between density and order, such as the Vicsek-model with topological interactions, 
Refs.~\cite{chou_12,ginelli_10a,peshkov_12_b}, do not show this instability.
In this Appendix, the mathematical details of the stability analysis for the standard VM will be given.

Density $\rho$ and momentum density $\vec{w}$ are expanded around a homogeneous ordered state, $\rho_0$, $\vec{w}_0=w_0\,\NX$ as
\begin{eqnarray}
\nonumber
\rho({\bf x},t)&=&\rho_0+\delta\rho\,{\rm e}^{i({\bf k}\cdot {\bf x}-\omega t)} \\
\nonumber         
\vec{w}({\bf x},t)&=&\vec{w}_0+(\TX\,\delta u+\NX\,\delta v)\,{\rm e}^{i({\bf k}\cdot {\bf x}-\omega t)} \\
\label{LINSTAB1}
{\bf k}&=&k_n\,\NX+k_t\,\TX\,,
\end{eqnarray}
where $\NX$ is the unit vector in the normal direction, defined by the direction of the unperturbed flying direction;
$\TX$ is the unit vector in the transverse direction with $\NX\cdot\TX=0$.
Consequently, $k_n$ and $k_t$ are the normal and the transverse components of the wave vector, respectively.
$\delta\rho$ is the amplitude of the density perturbation, $\delta v$ and $\delta u$ are the amplitudes
of the normal and transverse perturbations of the momentum, respectively.
$w_0$, the amplitude of the unperturbed momentum is given by
\begin{equation}
w_0=\sqrt{{1-\lambda\over q_3}}
\end{equation}
and is nonzero in the ordered phase where $\lambda>1$ and $q_3<0$.
Inserting the Ansatz, Eq. (\ref{LINSTAB1}), into the macroscopic equations, Eq. (\ref{FINAL_RHO}) and (\ref{FINAL_EVOLUTION_MOMENTUM})
and neglecting terms of higher than linear order in the perturbations
gives three coupled equations for $\delta \rho$, $\delta v$ and $\delta u$, which can be written in form of a homogeneous
system of linear equations,
$M\cdot \bf F=0$ 
where $F$ is a vector with ${\bf F}^T=(\delta\rho,\delta v,\delta u)$ and $M$ is a nonsymmetric $3\times 3$ matrix given by
\begin{eqnarray}
\nonumber
M_{11}&=&-\omega\\
\nonumber
M_{12}&=&k_n\\
\nonumber
M_{13}&=&k_t\\
\nonumber
M_{21}&=&ih_2k^2 k_n+h_4w_0 k^2-ih_3'k_n w_0^2-i\alpha k_n+q_2w_0(k_t^2-k_n^2) \\
\nonumber
         & & +i q_4 w_0^2 k_n+i k_3 w_0^2 k_n+q_3' w_0^3+\lambda' w_0^2\\
\nonumber
M_{22}&=&i\omega+h_1k^2-i(2h_3-q_1)w_0 k_n+3q_3w_0^2+\lambda-1\\
\nonumber
M_{23}&=&-i(2h_3+q_1)w_0 k_t \\
\nonumber
M_{31}&=&i h_2 k^2 k_t+ih_3' w_0^2 k_t-i\alpha k_t-2 q_2 k_n k_t w_0+iq_4 w_0^2 k_t-i k_3 w_0^2 k_t\\  
\nonumber
M_{32}&=&i(2h_3+q_1)w_0 k_t\\
M_{33}&=&i\omega+h_1k^2-i(2h_3-q_1) w_0 k_n+q_3 w_0^2+\lambda-1
\end{eqnarray}
Setting the determinant of this matrix to zero, gives the three branches of the dispersion relation $\omega(k)$, which
were obtained using Mathematica. The real parts of two of these branches are always negative but one of the branches
shows a long wave length instability in a small window $\eta_L<\eta<\eta_C$ below the threshold to collective (homogeneous) motion. In this window, the real part of omega is positive for wavenumbers $0\leq k \leq k_{max}$. More details on these results can be found in Refs. \cite{ihle_11,chou_12,ihle_15_a}.


\begin{thebibliography}{99} 

\bibitem{animal_flocks}
I.D. Couzin et al.,{\em Effective leadership and decision-making in animal groups on the move}, 
Nature {\bf 433}, 513 (2005).

\bibitem{BenJacob_97}
E. Ben-Jacob {\em et al.},{\em Chemomodulation of cellular movement, collective formation of vortices by swarming bacteria, and colonial development}, Physica A {\bf 238}, 181 (1997).

\bibitem{nano_rods}
Y.G. Tao and R. Kapral, {\em Swimming upstream: self-propelled nanodimer motors in a flow},
Soft Matter {\bf 6}, 756 (2010).

\bibitem{nedelec_02}
F. Nedelec, {\em Computer simulations reveal motor properties generating stable antiparallel microtubule interactions},
J. Cell Biology {\bf 158}, 1005 (2002).

\bibitem{actin_net}
J. F. Joanny {\em et al.},{\em Hydrodynamic theory for multi-component active polar gels},
New J. Phys. {\bf 9} 422 (2007). 

\bibitem{vicsek_12}
 T. Vicsek and A. Zafeiris,
 {\em Collective motion},
 Phys. Rep. {\bf 517} 71 (2012).

\bibitem{marchetti_13}
 M.C. Marchetti {\em et al.},
 {\em Hydrodynamics of soft active matter},
 Rev. Mod. Phys. {\bf 85} 1143 (2013).

\bibitem{toner_95}
J. Toner, Y. Tu,
{\em Long-Range Order in a Two-Dimensional Dynamical
XY Model: How Birds Fly together},
Phys. Rev. Lett. {\bf 75}, 4326 (1995).


\bibitem{toner_98}
J. Toner and Y. Tu,
{\em Flocks, herds, and schools: A quantitative theory of flocking},
Phys. Rev. E {\bf 58}, 4828 (1998).

\bibitem{vicsek_95}
T. Vicsek et al.,
{\em Novel type of phase transition in a system of self-driven particles},
Phys. Rev. Lett. {\bf 75}, 1226 (1995).

\bibitem{czirok_97}
A. Czirok, H.E. Stanley, T. Vicsek,
{\em Spontaneously ordered motion of self-propelled particles},
J. Phys. A: Math. Gen. {\bf 30}, 1375 (1997).

\bibitem{nagy_07}
M. Nagy, I. Daruka, T. Vicsek, 
{\em New aspects of the continuous phase transition in the scalar noise model (SNM) 
of collective motion},
Physica A {\bf 373}, 445 (2007).

\bibitem{ihle_11}
T. Ihle, 
{\em Kinetic theory of flocking: Derivation of hydrodynamic equations},
Phys. Rev. E {\bf 83},030901 (2011).

\bibitem{ihle_13}
T. Ihle, 
{\em Invasion-wave-induced first-order phase transition in systems of active particles},
Phys. Rev. E {\bf 88}, 040303 (2013).

\bibitem{ihle_14_a}
T. Ihle,
{\em Towards a quantitative kinetic theory of polar active matter},
Eur. Phys. J. Special Topics {\bf 223}, 1293 (2014);

\bibitem{ihle_15_a}
T. Ihle,
{\em 
Large density expansion of a hydrodynamic theory for self-propelled particles},
Eur. Phys. J. Special Topics {\bf 224}, 1303 (2015);

\bibitem{bertin_06}
E. Bertin, M. Droz, and G. Gr{\'e}goire,
{\em Boltzmann and hydrodynamic description for self-propelled particles},
Phys. Rev. E {\bf 74}, 022101 (2006).

\bibitem{bertin_09}
E. Bertin, M. Droz, and G. Gr{\'e}goire,
{\em Hydrodynamic equations for self-propelled particles: microscopic derivation and 
stability analysis},
J. Phys. A {\bf 42}, 445001 (2009).

\bibitem{baskaran_08b}
A. Baskaran, M. C. Marchetti,
{\em Hydrodynamics of self-propelled hard rods},
Phys. Rev. E {bf 77}, 011920 (2008).

\bibitem{grossmann_13}
R. Gro{\ss}mann, L. Schimansky-Geier, P. Romanczuk,
{\em Self-propelled particles with selective attraction-repulsion interaction: 
from microscopic 
dynamics to coarse-grained theories},
New J. Phys. {\bf 15}, 085014 (2013).

\bibitem{chepizhko_14}
O. Chepizhko, V. Kulinskii,
{\em The hydrodynamic description for the system of self-propelled particles: Ideal Vicsek fluid},
Physica A {\bf 415}, 493 (2014).

\bibitem{peshkov_12}
A. Peshkov, et al., 
{\em Nonlinear field equations for aligning self-propelled rods},
Phys. Rev. Lett. {\bf 109}, 268701 (2012).

\bibitem{farrell_12}
F.D.C. Farrell et al.,
{\em Pattern formation in self-propelled particles with density-dependent motility},
Phys. Rev. Lett. {\bf 108}, 248101 (2012).

\bibitem{peshkov_14_b}
A. Peshkov, E. Bertin, F. Ginelli, H. Chat{\'e}, 
{\em Boltzmann-Ginzburg-Landau approach for continuous 
descriptions of generic Vicsek-like models},
Eur. Phys. J Special Topics {\em 223}, 1315 (2014).

%---- START DEBATE ----------

\bibitem{ihle_14_b}
T. Ihle,
{\em 
Discussion  on  Peshkov  et  al.,  ``Boltzmann-
Ginzburg-Landau  approach  for  continuous
descriptions of generic Vicsek-like models''},
Eur. Phys. J. Special Topics {\bf 223}, 1427 (2014).

\bibitem{bertin_14_a}
E. Bertin et al.,
{\em Comment  on  Ihle,  ``Towards  a  quantitative
kinetic theory of polar active matter''},
Eur. Phys. J. Special Topics {\bf 223}, 1419 (2014).

\bibitem{ihle_14_c}
T. Ihle,
{\em 
Reply to comment on ``Towards a quantitative
kinetic  theory  of  polar  active  matter''  by
Bertin et al.''},
Eur. Phys. J. Special Topics {\bf 223}, 1423 (2014).

%---- END DEBATE ----------

\bibitem{BGK_54}
P. L. Bhatnagar, E. P. Gross, and M. Krook, 
{\em A Model for Collision Processes in Gases. I. 
Small Amplitude Processes in Charged and Neutral One-Component Systems},
Phys. Rev. {\bf 94}, 511 (1954).

\bibitem{klimon_67_74}
Yu. L. Klimontovich, {\em The Statistical Theory of Nonequilibrium Processes
in a Plasma}, (Pergamon, London, 1967);
Sov. Phys. Usp. {\bf 16}, 512 (1974).

\bibitem{ernst_81}
M.H. Ernst, E.G.D. Cohen,
{\em Nonequilibrium fluctuations in $\mu$ space},
J. Stat. Phys. {\bf 25}, 153 (1981).

\bibitem{nicholson_83}
D. R. Nicholson, {\em Introduction to Plasma Theory}, 
John Wiley \& Sons, New York, 1983.

\bibitem{chou_15}
Y.-L. Chou and T. Ihle,
{\em Active matter beyond mean-field: Ring-kinetic theory for self-propelled particles},
Phys. Rev. E {\bf 91}, 022103 (2015).

\bibitem{archer_04}
A.J. Archer and M. Rauscher, {\em Dynamical density functional theory for interacting 
Brownian particles: stochastic or deterministic?},
J. Phys A: Math. Gen. {\bf 37}, 9325 (2004).

\bibitem{thueroff_13}
%critical assessment
F. Th{\"u}roff, C.A. Weber, E. Frey,
{\em Critical Assessment of the Boltzmann Approach to Active Systems},
Phys. Rev. Lett. {\bf 111}, 190601 (2013).

\bibitem{hanke_13}
%active colloids
T. Hanke, C.A. Weber, E. Frey,
{\em 
Understanding collective dynamics of soft active colloids by binary scattering},
Phys. Rev. E {\bf 88}, 052309 (2013).

\bibitem{FOOT1}
A crucial step in these derivations is at what time and for which coordinates the factorization assumption, 
Eq. (\ref{PROD_ANSATZ}), is applied. This choice distinguishes between ``before'' and ``after'' a collision and
leads to the irreversibility of the final Boltzmann equation.

\bibitem{grad_58}
H. Grad,
{\em Principles of the kinetic theory of gases},
in ``Thermodynamics of gases'', encyclopedia of physics, Vol. 12, pp 205-294,
ed. by S. Fl{\"u}gge, Springer, 1958.

\bibitem{kreuzer_81}
H.J. Kreuzer,
{\em Nonequilibrium thermodynamics and its statistical foundations},
Oxford and New York, Clarendon Press, 1981.

\bibitem{cercignani_88}
C. Cercignani,
{\em The Boltzmann equation and its applications},
Vol. 67 of series ``Applied Mathematical Sciences'', Springer, 1988.

\bibitem{enskog_21}
D. Enskog, K. Svenska Vetenskaps Akademiens Handl., 1921, 63 no. 4; 
english transl. in {\em Kinetic Theory}, ed. S. Brush, Pergamon Press, London, New York, 1972, vol. 3. 

\bibitem{chapman_52}
S. Chapman and T. G. Cowling, {\em The Mathematical Theory of Non-Uniform Gases},
Cambridge University Press, Cambridge, 1952.

\bibitem{hirschfelder_54}
J. O. Hirschfelder, C. F. Curtiss and R. B. Bird,
{\em Molecular Theory
of Gases and Liquids},
John Wiley \& Sons, New York, 1954.

\bibitem{mcquarrie_76}
D.  A.  McQuarrie,
{\em Statistical  Mechanics},
Harper  \&  Row,
New York, 1976.

\bibitem{frisch_87}
U. Frisch {\em et al.},
{\em Lattice gas hydrodynamics in two and three dimensions},
Complex Systems {\bf 1}, 649 (1987).

\bibitem{chen_98}
S. Chen and G.D. Doolen, {\em Lattice-Boltzmann method for fluid flow},
Annu. Rev. Fluid Mech. {\bf 30}, 329 (1998).

\bibitem{ihle_00}
T. Ihle and D.M. Kroll, {\em Thermal lattice-Boltzmann method for non-ideal gases
with potential energy},
Comp. Phys. Comm. {\bf 129}, 1 (2000).

\bibitem{ihle_09}
T. Ihle, 
{\em Chapman-Enskog expansion for multi-particle collision models},
Phys. Chem. Chem. Phys. {\bf 11}, 9667 (2009).

\bibitem{mcnamara_93}
G. McNamara and B. Alder, {\em Analysis of the lattice Boltzmann treatment of hydrodynamics},
Physica A {\bf 194}, 218 (1993).

\bibitem{FOOT2}
Note, this does not correspond to the dilute limit, which is rather characterized by $M\ll 1$. 
Considering genuine multiparticle collisions, that is $M>1$,
is relevant for kinetic theories \cite{chou_12} of systems like flocks of starlings \cite{ballerini_08} 
where a bird was found to interact with 6 to 7 neighbors at once.

\bibitem{ballerini_08}
M. Ballerini et al.,
{\em Interaction ruling animal collective behavior depends on topological rather than metric distance: Evidence from a field study},
PNAS {\bf 105}, 1232 (2008).

\bibitem{hanley_72}
H.J.M. Hanley, R.D. McCarty, E.G.D. Cohen,
{\em Analysis of the transport coefficients of simple dense fluids: application of the modified Enskog theory},
Physica A {\bf 60}, 322 (1972).

\bibitem{gompper_08}
G. Gompper, T. Ihle, D.M. Kroll, R.G. Winkler,
{\em Multi-Particle Collision Dynamics: a particle-based mesoscale simulation approach to the 
hydrodynamics of complex fluids},
Adv. Polym. Sci. {\bf 221}, 1 (2008).

\bibitem{alexander_98}
F.J. Alexander, A.L. Garcia, B.J. Alder,
{\em Cell size dependence of transport coefficients in stochastic particle dynamics},
Phys. Fluids {\bf 10}, 1540 (1998).



\bibitem{chou_12}
Y. L. Chou, R. Wolfe, and T. Ihle,
{\em Kinetic theory for systems of self-propelled particles with metric-free interactions},
 Phys. Rev. E {\bf 86}, 021120 (2012).


\bibitem{ihle_unpub}
T. Ihle, in preparation.

\bibitem{aranson_05}
I.S. Aranson, L.S. Tsimring,
{\em Pattern formation of microtubules and motors: 
Inelastic interaction of polar rods},
Phys. Rev. E {\bf 71}, 050901 (2005).

\bibitem{tuzel_03} 
E. T\" uzel, M. Strauss, T. Ihle and D.M. Kroll, 
{\em Transport coefficients for stochastic rotation dynamics in three dimensions},
Phys. Rev.
E {\bf 68}, 036701 (2003).

\bibitem{ihle_05}
T. Ihle, E.T\"uzel, and D.M. Kroll,
{\em Equilibrium calculation of transport coefficients for a fluid-particle model},
Phys. Rev.
E {\bf 72}, 046707 (2005).

\bibitem{ginelli_10a}
F. Ginelli, H. Chat{\'e},
{\em Relevance of Metric-Free Interactions in Flocking Phenomena},
Phys. Rev. Lett. {\bf 105}, 168103 (2010).

\bibitem{peshkov_12_b}
A. Peshkov et al.,
{\em Continuous Theory of Active Matter Systems with Metric-Free Interactions},
Phys. Rev. Lett.
{\bf 109}, 098101 (2012).

\end{thebibliography}
\end{document}